\documentclass[aps,prb,twocolumn, superscriptaddress, showpacs,draft]{revtex4-1}
\usepackage{amsmath,amsthm}
\usepackage{amssymb,amsfonts}
\usepackage{graphicx}
\usepackage{stmaryrd}
\usepackage{lmodern}
\usepackage{slantsc}
\usepackage{scalefnt}
\usepackage{subfigure}
\usepackage{dsfont}
\usepackage{xspace}
\usepackage{boxedminipage}
\usepackage{ifpdf}
\usepackage{pgffor}
\usepackage{xifthen}
\usepackage{tensor}
\usepackage{array}
\usepackage{tikz}
\usetikzlibrary{shapes.geometric}
\usepackage{color}
\usetikzlibrary{arrows,matrix,calc,scopes,decorations.markings}
\usepackage{tikz-cd}
\allowdisplaybreaks[1]
\usepackage[mathcal]{euscript}
\newtheorem{theorem}{Theorem}

\newtheorem{lemma}[theorem]{Lemma}

\newtheorem*{ansatz*}{Ansatz}
\newcommand{\be}{\begin{equation}}
\newcommand{\ee}{\end{equation}}
\newcommand{\bse}{\begin{subequations}}
\newcommand{\ese}{\end{subequations}}
\newcommand{\ket}[1]{|{#1}\rangle}

\newcommand{\Z}{\mathbb{Z}}
\newcommand{\ii}{\mathrm{i}}

\newcommand{\Hil}{\mathcal{H}}

\newcommand{\F}{\mathfrak{F}}
\newcommand{\T}{\mathcal{T}}
\newcommand{\A}{\mathcal{A}}

\newcommand{\C}{\mathcal{C}}
\newcommand{\B}{\mathcal{B}}
\newcommand{\FF}{\mathcal{F}}
\newcommand{\bpm}{\begin{pmatrix}}
\newcommand{\epm}{\end{pmatrix}}
\newcommand{\bmm}{\begin{matrix}}
\newcommand{\emm}{\end{matrix}}

\newcommand{\x}{\times}
\newcommand{\ox}{\otimes}
\newcommand{\bx}{\boxtimes}
\newcommand{\orbx}{\overrightarrow{\bx}}

\newcommand{\diag}{\mathrm{Diag}}
\newcommand{\rep}{\mathrm{Rep}}
\newcommand{\Hom}{\mathrm{Hom}}
\newcommand{\id}{\mathrm{id}}

\newenvironment{smallarray}[1]
 {\null\,\vcenter\bgroup\scriptsize
  \arraycolsep=.13885em
  \hbox\bgroup$\array{@{}#1@{}}}
 {\endarray$\egroup\egroup\,\null}

\newcommand{\vlc}[6]{
\bigl[\hspace{-0.2em}
\begin{smallarray}{cc|c}
    #1 & #2 & #3\\ [0.2em]
    #4 & #5 & #6
\end{smallarray}\hspace{-0.2em}\bigr] }

\newcommand{\Fm}[7][0]{
\ifthenelse{\NOT \equal{#1}{0}}{
\left[F^{#2#3#4}_{#5}\right]_{#6#7}
}
{\bigl[F^{#2#3}_{#4#5}\bigr]^{#6}_{#7}}
}

\makeatletter
\newcommand*{\Relbarfill@}{\arrowfill@\Relbar\Relbar\Relbar}
\newcommand*{\xeq}[2][]{\ext@arrow 0055\Relbarfill@{#1}{#2}}
\makeatother

\newcommand{\gdw}[2]{
\definecolor{c0000d5}{RGB}{0,0,213}
\definecolor{c70a608}{RGB}{112,166,8}

\begin{tikzpicture}[y=0.80pt, x=0.80pt, yscale=-1.000000, xscale=1.000000, inner sep=0pt, outer sep=0pt]
  \path[draw=black,fill=c0000d5,opacity=0.374,line join=miter,line cap=butt,even odd rule,line width=0.595pt,rounded corners=0.0000cm] (49.8716,32.2338) rectangle (104.9241,112.4906);
  \node[inner sep=0,outer sep=0] at (78,72) {\scalebox{1}{$#1$}};
  \path[draw=black,fill=c70a608,opacity=0.374,line join=miter,line cap=butt,even odd rule,line width=0.595pt,rounded corners=0.0000cm] (105.0000,32.3622) rectangle (160.0525,112.6190);
  \node[inner sep=0,outer sep=0] at (132,72) {\scalebox{1}{$#2$}};
\end{tikzpicture}
 }
 
\newcommand{\gdwFold}[2]{
\definecolor{c70a608}{RGB}{112,166,8}
\definecolor{c0000d5}{RGB}{0,0,213}
\definecolor{00000ff}{RGB}{0,0,0}

\begin{tikzpicture}[y=0.80pt, x=0.80pt, yscale=-1.000000, xscale=1.000000, inner sep=0pt, outer sep=0pt]   

  \path[cm={{0.98995,0.14141,0.0,1.0,(0.0,0.0)}},draw=black,fill=c70a608,opacity=0.304,line join=miter,line cap=butt,even odd rule,line width=0.573pt,rounded corners=0.0000cm] (188.2203,-1.6757) rectangle (239.3506,78.6108);

  \path[draw=black,fill=c0000d5,opacity=0.4,line join=miter,line cap=butt,even odd rule,line width=0.595pt,rounded corners=0.0000cm] (181.8716,32.2338) rectangle (236.9241,112.4906);

  \node[inner sep=0,outer sep=0] at (208,72) {\scalebox{1}{$#1$}};
  
  \path[draw=black,fill=00000ff,opacity=0.2,line join=miter,line cap=butt,miter limit=4.00,even odd rule,line width=0.476pt,rounded corners=0.0000cm] (236.9377,32.1818) rectangle (277.5986,112.5427);
  \node[inner sep=0,outer sep=0] at (256,72) {\scalebox{1}{$#2$}};
\end{tikzpicture}
}

\newcommand{\gdwTwo}[5]{

\definecolor{c0000d5}{RGB}{0,0,213}
\definecolor{c70a608}{RGB}{112,166,8}

\begin{tikzpicture}[y=0.80pt, x=0.80pt, yscale=-1.000000, xscale=1.000000, inner sep=0pt, outer sep=0pt]

  \path[draw=black,fill=c0000d5,opacity=0.564,line join=miter,line cap=butt,even odd rule,line width=0.595pt,rounded corners=0.0000cm] (49.9475,27.3622) rectangle (105.0000,107.6190);
  \node[inner sep=0,outer sep=0] at (78,68) {\scalebox{1}{$#1$}};   
  \path[draw=black,fill=c0000d5,opacity=0.374,line join=miter,line cap=butt,even odd rule,line width=0.595pt,rounded corners=0.0000cm] (105.0000,27.3622) rectangle (160.0525,107.6190);
  \node[inner sep=0,outer sep=0] at (133,68) {\scalebox{1}{$#2$}}; 
  \node[inner sep=0,outer sep=0] at (105,115) {\scalebox{1}{$#4$}};   
  \path[draw=black,fill=c70a608,opacity=0.374,line join=miter,line cap=butt,even odd rule,line width=0.595pt,rounded corners=0.0000cm] (160.0000,27.3622) rectangle (215.0525,107.6190);
  \node[inner sep=0,outer sep=0] at (188,68) {\scalebox{1}{$#3$}}; 
  \node[inner sep=0,outer sep=0] at (160,115) {\scalebox{1}{$#5$}};
\end{tikzpicture}
}

\newcommand{\gdwNo}[3]{
\definecolor{c0000d5}{RGB}{0,0,213}
\definecolor{c70a608}{RGB}{112,166,8}

\begin{tikzpicture}[y=0.80pt, x=0.80pt, yscale=-1.000000, xscale=1.000000, inner sep=0pt, outer sep=0pt]
  \path[draw=black,fill=c0000d5,opacity=0.374,line join=miter,line cap=butt,even odd rule,line width=0.595pt,rounded corners=0.0000cm] (49.8716,32.2338) rectangle (104.9241,112.4906);
  \node[inner sep=0,outer sep=0] at (78,72) {\scalebox{1}{$#1$}};
  \path[draw=black,fill=c70a608,opacity=0.374,line join=miter,line cap=butt,even odd rule,line width=0.595pt,rounded corners=0.0000cm] (105.0000,32.3622) rectangle (160.0525,112.6190);
  \node[inner sep=0,outer sep=0] at (132,72) {\scalebox{1}{$#2$}};
  \node[inner sep=0,outer sep=0] at (105,120) {\scalebox{1}{$#3$}};
\end{tikzpicture}
 }

\begin{document}

\title{Fermion Condensation and Gapped Domain Walls in Topological Orders}

\author{Yidun Wan}
\email{ywan@perimeterinstitute.ca}
\affiliation{Department of Physics and Center for Field Theory and Particle Physics, Fudan University, Shanghai 200433, China}
\affiliation{Perimeter Institute for Theoretical Physics, Waterloo, ON N2L 2Y5, Canada}
\author{Chenjie Wang}
\email{cwang@perimeterinstitute.ca}
\affiliation{Perimeter Institute for Theoretical Physics, Waterloo, ON N2L 2Y5, Canada}

\date{\today}

\begin{abstract}
We propose the concept of fermion condensation in bosonic topological orders in two spatial dimensions. 
Fermion condensation can be realized as gapped domain walls between bosonic and fermionic topological orders, which are thought of as a real-space phase transitions from bosonic to fermionic topological orders. This generalizes the previous idea of understanding boson condensation as gapped domain walls between bosonic topological orders. We show that generic fermion condensation obeys a Hierarchy Principle by which it can be decomposed into a boson condensation followed by a minimal fermion condensation, which involves a single self-fermion that is its own anti-particle and has unit quantum dimension. We then develop the rules of minimal fermion condensation, which together with the known rules of boson condensation, provides a full set of rules of fermion condensation. Our studies point to an exact mapping between the Hilbert spaces of a bosonic topological order and a fermionic topological order that share a gapped domain wall.

\end{abstract}
\pacs{11.15.-q, 71.10.-w, 05.30.Pr, 71.10.Hf, 02.10.Kn, 02.20.Uw}
\maketitle

\section{Introduction}\label{sec:intro}
A gapped quantum matter phase with intrinsic topological order has topologically protected ground state degeneracy and anyon excitations\cite{Wen1990a,Wen1991} on which quantum computation may be realized via anyon braiding, which is robust against errors due to local perturbation\cite{Kitaev2003a,Kitaev2006}. Topological orders are also believed to have a correspondence with topological field theories, which effectively describe the ground states of topological orders. Towards the applications of topological orders, practical or theoretical, it is of paramount importance to understand the classification of topological orders and phase transitions between topological orders. While the classification of topological order has been studied for a long time, phase transitions between topological orders are much less understood.

One kind of phase transitions between topological orders that have been studied for a while is called {\it anyon condensation}.
The idea of anyon condensation can be traced back to anyon superconductivity, first proposed by Laughlin\cite{Laughlin1988,Laughlin1988a} and followed by Wilczek and others\cite{Fetter1989,CHEN1989}, which was studied to possibly account for high-$T_c$ superconductivity. Recently, Wen \textit{et al} propose a system providing the right energy regime for anyon gases. In these proposals, a collection of anyons may form a boson and then condense, in a fashion similar to Cooper pair condensation. More recently, Bais \textit{et al} proposed a set of empirical rules of condensing self-bosons that have nontrivial braiding statistics with some other anyons in a bosonic topological order (bTO).\cite{Bais2002,Bais2009,Bais2009a,Bais2012}  After condensation, the condensed self-bosons result in a new vacuum, and the original bTO undergoes a phase transition to a new bTO. It was later found\cite{HungWan2015a} that mathematically, the set of condensed self-bosons corresponds to a special type of Frobenius algebras\cite{Fuchs2002,Fuchs2002a,Fuchs2004a,Kirillov2002,Fuchs2013}, which are  objects in the unitary modular tensor categories (UMTCs) that describe the bTO.

Previous studies on anyon condensation are restricted to condensing self-bosons in a bTO. A particularly interesting question is: Is it possible to condense self-fermions, which have nontrivial braiding statistics with some other anyons in the system? At a first glance, fermion condensation might be counterintuitive; however, in this work, we propose a physical context in which fermion condensation is perfectly reasonable and develop a theory of fermion condensation.

\begin{figure}[b]
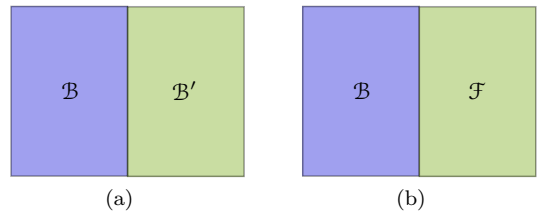

\subfigure[]{\label{subfig:Bgdw1}
\gdw{\B}{\B'}
}
\subfigure[]{\label{subfig:Fgdw1}
\gdw{\B}{\FF}
}
\caption{Gapped domain walls between two bTOs (a) and between a bTO and an fTO (b).} \label{fig1}
\end{figure}

To make sense of fermion condensation, let us take a step back and discuss the following two viewpoints on the boson condensation proposed by Bais \textit{et al}. On the one hand, let us imagine tuning some parameter in the Hamiltonian of the bTO, such that the energy gap closes and reopens, giving rises to a new bTO. For example, one may consider a $G$ discrete gauge theory, which holds a topological order described by the quantum double $D[G]$. We can drive a Higgs transition by condensing certain charge excitations such that the gauge group $G$ is broken down to $H\subset G$, leading to a new bTO $D[H]$. However,  the physical mechanism behind such ``parameter space'' phase transition for general bTOs is still unclear.

On the other hand, which may be physically more transparent, let us consider gapped domain walls (GDW) between bTOs (see Fig.~\ref{fig1}(a))---phase transitions in ``real space''. Imagine a bTO $\B$. After condensing certain self-bosons, it becomes $\B'$. In real space, this boson condensation can be viewed as a special type of GDWs between $\B$ and $\B'$, where every anyon in $\B'$ can penetrate the GDW and transform into some anyon in $\B$, rather, some anyons in $\B$ may never be able to penetrate the GDW (i.e., they are confined in $\B'$ in the sense of parameter space phase transition). The self-bosons that condense in $\B$ transform into the trivial boson $1\in \B'$, when they cross the GWD. The connection between GDWs in bTOs and self-boson condensation has been thoroughly studied recently.\cite{Kitaev2012,Gaiotto2012,Kong2013,Gu2014a,HungWan2014,Lan2014,Gaiotto2014, HungWan2015a} It has been found that GDWs between bosonic topological orders can be classified by anyon condensation\cite{Kong2013,HungWan2015a}.

That being said, we are now ready to explore self-fermion condensation in bTOs. By self-fermion condensation, we mean some kind of ``condensation transition'' through which certain self-fermions in a bTO  become {\it local excitations} after the condensation. If we imagine such a phase transition in parameter space, one finds that the original {\it bosonic} system turns into a {\it fermionic} system after the condensation---because, with the current knowledge of topological orders, only fermionic systems, where the fundamental degrees of freedom are fermions, support local fermionic excitations. In bTOs, all self-fermions are nonlocal. It is however still mysterious how can one turn a bosonic system into a fermionic system through certain phase transition if without involving, say, for example, a background layer of fermionic system.

On the other hand, the GDW picture makes perfect sense for discussing about self-fermion condensation. We just need to consider a GDW between a bTO and a fermionic topological order (fTO)\cite{Gu2014b,Lan2015,Bhardwaj2016}, whose fundamental degrees of freedom are fermions (Fig.~\ref{fig1}(b)). From the GDW picture of self-boson condensation, we expect that a self-fermion condensation has the following properties: (1) all excitation in $\FF$ can pass through the wall and become some excitations in $\B$; (2) the condensed self-fermions in $\B$ becomes local fermions in $\FF$ when they pass through the wall; (3) certain anyons in $\B$ cannot pass through the wall. In general, we expect that self-bosons and self-fermions can condense simultaneously, and we have (4) the condensed self-bosons in $\B$ becomes the trivial boson in $\FF$. Of course, these properties only give a general intuitive picture of self-fermion condensation.

The goal of this paper is to establish more precise rules on self-fermion condensation, or simply put, fermion condensation, with the GDW picture in mind. To this end, we propose a physically and mathematically reasonable ansatz for fermion condensation. Based on this ansatz, we prove a Hierarchy Principle of fermion condensation in bTOs, in the sense that any well-defined fermion condensation,  however complicated, can be decomposed into two steps: First, we can collect all self-bosons in the condensate and condense them alone; Second, what remains is only to condense a single self-fermion, which we call {\it minimal} fermion condensation. We then find the rules of minimal fermion condensation, which, combined with the known rules of boson condensation, give rise to a full set of rules of performing generic fermion condensation in bTOs.

With these rules, we study properties of fermion condensation. In particular, we reach a formula between the total quantum dimension of the parent bTO and that of its child bTO due to certain fermion condensation. The previously known formula for boson condensation\cite{HungWan2015a} appears to be a special case of our new formula. In addition, we find that there is an equivalence between fermion condensation in bTOs and boson condensation in fTOs. This equivalence in turn corroborates our rules of fermion condensation.  We work out explicit examples to illustrate the results described above. Finally in the discussion section, we take an outlook into certain future directions.

We believe that fermion condensation can also be described by Frobenius algebras that are of a type different from the type of Frobenius algebras characterizing boson condensation. Nevertheless, in this work, we would not dwell on the abstract math of fermion condensation but instead focus on the physical content and consequences of fermion condensation by minimizing the mathematics. While we shall report the results of our ongoing research of the full mathematics of fermion condensation elsewhere, in the appendix we would briefly review the type of Frobenius algebras describing boson condensation and show what constraints are relaxed on bosonic Frobenius algebra to obtain a fermionic Frobenius algebra.

The study of fermion condensation can in turn help us understand fTOs. Although the most common condensed matter systems---electron systems---are fermionic, fTOs are less understood than bTOs. Currently, a mathematically rigorous and complete description of fTOs is yet unavailable. It is therefore worthwhile of understanding fTOs indirectly from other perspectives. With fermion condensation, we can construct new fTOs based on bTOs, and thereby extract a proper description of fTOs from the data of bTOs.

As a warning of terminology abusing: Throughout the paper, we refer to self-bosons (fermions) simply by bosons (fermions) wherever no confusion would arise; We often do not differentiate an anyon type from an anyon if the context is clear, otherwise we refer to an anyon type as a topological sector; We may also use bTOs and unitary modular tensor categories interchangeably.

It is worth of note that simple-current fermion condensation, which is equivalent to our minimal fermion condensation, was briefly studied in Refs\cite{Gaiotto2015}. During the preparation of the manuscript, a more detailed study of simple current fermion condensation was done by Gaiotto \textit{et al}\cite{Bhardwaj2016}. In this work, we consider general fermion condensation, and propose a realization through GDWs between bTOs and fTOs.

\section{Bosonic and Fermionic topological orders}\label{sec:topoOrder}
We begin with a brief review of the basics of bTOs and fTOs. A \textit{bTO} is a 2D quantum many-body system with an energy gap, and the excitations are generally {\it anyons}. The microscopic degrees of freedom underlying a bTO are bosons, such that the vacuum state (also called the trivial boson) is a true boson and unique, which has trivial mutual statistics with every anyon in the system. Anyons in a bTO are believed to be fully characterized by a unitary modular tensor category (UMTC) $\B$, where the modularity refers to that the trivial boson is the only topological sector with trivial mutual statistics with all topological sectors of the topological order. A bTO $\B$ is composed of the following topological data. First, a finite set of anyons $\{1, a,b,\dots\}$ equipped with quantum dimensions $\{1,d_a,d_b,\dots\}$, where $d_a\geq 1$ are real numbers. The unitarity demands the positivity of quantum dimensions. Here, 1 denotes the trivial anyon and $d_1=1$. The dimension of the UMTC $\B$ or the total quantum dimension of the bTO $\B$ is defined by $D_\B=\sqrt{\sum_{a\in\B}d_a^2}$, where the sum includes the trivial boson $1$ as well. Second, the fusion interaction between the anyons, namely $a\x b=\sum_c N^c_{ab} c$, where the sum is over all anyons including the trivial anyon $1$, and $N^c_{ab}$ are nonnegative integers. The fusion is associative, and $1$ is the identity of fusion. Each anyon $a$ in $\B$ has an \textit{anti-anyon} $\bar a\in \B$, such that $1\in a\x\bar a$. The anti-anyon $\bar a$ of $a$ is unique, and the vacuum $1$ can only appear exactly once in $a\x\bar a$, i.e., $N^1_{a\bar a}\equiv 1$. Following Bais' convention, we also call this feature the unitarity condition\cite{Bais2009}, which should not be confused with the unitarity of UTMC. An anyon $a$ may be \textit{self-dual} in the sense that $a=\bar a$. Third, a modular $T$ matrix, $T=\diag\{1,\theta_a,\theta_b,\dots\}$ where $\theta_a=\exp(\ii 2\pi h_a)$ is the self-statistical angle of $a$ with $h_a$ the topological spin of $a$. That $\B$ is a unitary also implies that $\theta_a=\theta_{\bar a}$. Fourth, a modular $S$ matrix encoding the braiding between the anyons, whose matrix elements are
\be\label{eq:S}
S_{ab}=\frac{1}{D}\sum_c N^c_{ab}\frac{\theta_c}{\theta_a\theta_b}d_c,
\ee
where the total quantum dimension $D=\sqrt{\sum_a d_a^2}$. This definition of $S$-matrix implies that $S_{1a}=d_a/D$.

An \textit{fTO}, however, has not only a trivial boson but also a trivial fermion (also known as a transparent fermion) that has trivial mutual statistics with all topological sectors in the fTO. Hence, unlike a bTO, an fTO does not respect modularity. The existence of a trivial fermion is a result of the fermionic microscopic degrees of freedom underlying the topological order. The trivial fermion has unit quantum dimension too. Although an fTO is not described by a UMTC, it is still a fusion category; hence, its total quantum dimension takes the same definition as that of a bTO. A general categorical theory of fTOs is still unclear but certain features of fTOs, such as the non-modularity, are captured by what are known as premodular tensor categories\cite{Lan2015}. Nevertheless, several properties of fTO's are understood from physics point of view. In general, an fTO $\FF$ contains a finite set of anyons $\{1,1^f,a,a^f,b,b^f,\dots \}$, where 1 is the trivial boson and $1^f$ the trivial fermion. Topological sectors always come in pairs, $a$ and $a^f$,  with $a^f = a\times1^f$. The self-statistical angles of anyons in each pair satisfy $\theta_{a^f} = -\theta_a$, following the fact that $1^f$ is a fermion and is transparent.  An fTO that contains only $1$ and $1^f$ is regarded as a trivial fTO and denoted by $\FF_0=\{1,1^f\}$. A trivial fTO can be realized in gapped free fermion systems.  Note that for any fTO $\FF$, $\FF\supseteq \FF_0$.  The anti-anyon $\bar a$ is defined in the same way as in bTOs, such that $1\in a\times\bar a$. The anti-anyon is unique and the trivial boson $1$ occurs in $a\times\bar a$ only once.

Fermionic topological orders can be divided into two types, namely {\it primitive} fTOs and {\it non-primitive} ones\cite{Lan2015}. Primitive fTOs are those that cannot be obtained from stacking two other topological orders, while non-primitive ones can. There are two ways to obtain fTO's from stacking: (1) stack an fTO with a bTO and (2) stack two fTO's.  The simplest kind of non-primitive fTOs are those obtained by stacking a layer of bTO $\B$ with a layer of trivial fTO $\FF_0$. We denote this stacking operation by $\B\bx\FF_0$. It is shown that any Abelian fTO admits this kind of layer decomposition\cite{Lan2015}.  One can also stack two different fTOs $\FF_1$ and $\FF_2$ together by $\FF_1\bx_{\FF_0}\FF_2$, the symbol $\bx_{\FF_0}$ means that the $\FF_0\subset\FF_1$ and $\FF_0\subset\FF_2$, two copies of $\FF_0$, must be identified as a single $\FF_0$. Certainly, one can stack any finite number of TOs in this manner to construct more complicated TOs, bosonic or fermionic. As such, a \textit{primitive} fTO does not admit any nontrivial stacking structure, other than the trivial stacking $\FF=\FF \bx_{\FF_0} \FF_0$. The topological properties of the TOs obtained by such stacking operations can be straightforwardly obtained from those of individual layers. The anyons in some $\B\bx\FF$ are the pairs $(a,a')$ $\forall a\in\B,\ a'\in\FF$. The quantum dimensions read $d_{(a,a')}=d_ad_{a'}$. Topological spins are sums, $h_{(a,a')}=h_a+h_{a'}$. The modular $S$ matrix of $\B\bx\FF$ is the tensor product of that of $\B$ and that of $\FF$. We will bring up an important difference between primitive and non-primitive fTOs in Section \ref{subsec:otherGDW}.

Finally, we make comments on the trivial bTO $\B_0 = \{1\}$ and trivial fTO $\FF_0 = \{1,1^f\}$. The above discussions focus on the anyonic content of topological orders. Two topological orders with exactly the same anyonic content may still be distinct, in the sense that they cannot be smoothly deformed to each other without closing the energy gap. It is believed that to fully characterize a topological order, one needs an extra piece of data, the chiral central charge $c$ of the edges modes (some conformal field theory)  living on the boundary of the 2D topologically ordered system.  Therefore, there are various bTOs/fTOs with the same anyonic content but different values of $c$. It is believed that bTOs with $\B_0$ can be obtained by stacking the $E_8$ states\footnote{\uppercase{K}. Walker, Presentation:\textit{Codimension 1 defects, categorified group actions, and condensing fermions}, url: http://canyon23.net/math/talks/IPAM 201501b compressed.pdf} which has $c=8$, and fTO's with $\FF_0$ can be obtained by stacking $p + \ii p$ states\cite{Read2000}\footnote{\uppercase{A}. Kitaev, Presenation:\textit{Toward Topological Classification of Phases with Short-range Entanglement}, url: http://online.kitp.ucsb.edu/online/topomat11} which has $c=1/2$. In the case of gapped domain walls, $c$ is always the same on the two sides of the wall. So, we will not emphasize on $c$ any further in the sequel. In fact, we will and have already abused to certain extent the notions of a TO and its anyonic content. We will ignore the variety in $c$, and equate a TO with its anyonic content.

\section{From Boson condensation to fermion Condensation}\label{sec:ferCond}
In this section, we write down our ansatz for fermion condensation in bTOs, which consists of several necessary conditions for fermion condensation to satisfy. To motivate this ansatz, we retrospect to and quickly go through the basic idea of boson condensation and the GDWs between bTOs. The core notion arises there is the \textit{mutual locality} between condensed bosons. As to be seen, we can naturally extend boson condensation to fermion condensation and write down a reasonable ansatz.

Actually, it is plausible that one can generalize the idea to condensation of arbitrary anyons, which we would briefly touch upon in the discussion Section \ref{sec:Disc}.

\subsection{Boson Condensation and GDWs}\label{subsec:BCandGDW}
\begin{figure}[h!]
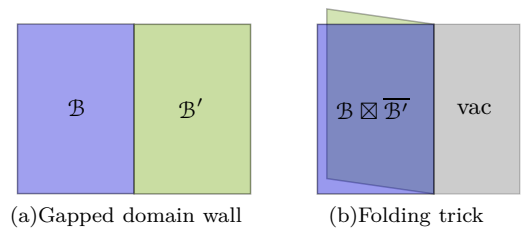

\subfigure[Gapped domain wall]{\label{subfig:Bgdw}
\gdw{\B}{\B'}
}
\subfigure[Folding trick]{\label{subfig:Bfold}
\gdwFold{\B\bx\overline{\B'}}{\rm{vac}}
}
\caption{(a) A bTO $\B$ and a $\B'$ connected by a GDW via condensing some bosons in $\B$. Via the folding trick along the GDW, this picture is equivalent to (b) A bTO $B\bx\overline{\B'}$ sharing a gapped boundary with the vacuum.}
\label{fig:BgdwAndFolding}
\end{figure}

Consider two (not necessarily different) bTOs $\B$ and $\B'$ that share a GDW in between (Fig. \ref{fig:BgdwAndFolding})\footnote{We consider only 1D GDWs in this paper. In general, one may think of 2D GDWs between $\B$ and $\B'$, which itself is a TO $\B''$ that shares 1D GDWs with both $\B$ and $\B'$.}. To understand the GDW between $\B$ and $\B'$, we have two equivalent ways. On the one hand, we can fold the system along the GDW and turn it into the configuration in Fig. \ref{subfig:Bfold}. The bTO $\B\bx\overline{\B'}$ is a simple stacking of $\B$ and $\B'$, where $\overline{\B'}$ means that the spins and braiding statistics should be a mirror reflection of those of $\B'$. We can visualize the GDW (or called a gapped boundary in this case) between $\B\bx\overline{\B'}$ and the vacuum as follows. The anyons in $\B\bx\overline{\B'}$ take the form $(a,b)$, where $a\in\B$ and $b\in\B'$. Imagine one creates an anyon $(a,b)$ in the bulk of $\B\bx\overline{\B'}$ and moves it to the gapped boundary. Since the boundary is gapped, certain such anyons would have to be destroyed at the boundary by local operators acting on the boundary and become part of the vacuum. In other words, they are condensed on the boundary. Condensing such anyons at the boundary results in a gapped boundary, in a way analogous to the Higgs mechanism or Cooper pair condensation. But not all anyons in $\B\bx\overline{\B'}$ can condense at the boundary and disappear into the vacuum. One may trade this picture of gapped boundary with a phase transition in certain parameter space, in which one can physically trigger a Higgs transition from $\B\bx\overline{\B'}$ to the vacuum by condensing those anyons that can disappear at the boundary. In this scenario, those anyons of $\B\bx\overline{\B'}$ that cannot disappear at the boundary would become confined through the phase transition and are not physical excitations in the vacuum. Condensability and confinement are tied to the notion of mutual locality to be discussed shortly.


On the other hand, one can unfold the gapped boundary picture in Fig. \ref{subfig:Bfold} back to the original picture of GDW in Fig. \ref{subfig:Bgdw}. This way, one can see that an anyon $(a,b)$ in $\B\bx\overline{\B'}$ that can disappear at the gapped boundary corresponds to two anyons, $a\in\B$ and $b\in\B'$, such that they can meet at the GDW between $\B$ and $\B'$ and disappear {\it simultaneously}. Alternatively, it can be viewed as that $a$ can cross the GDW and turn into $b$, and vice versa. Meanwhile, an anyon $(a,b)\in \B\bx\overline{\B'}$ that cannot disappear at the gapped boundary in the folded picture corresponds to the situation that there is no way (operator) to turn $a$ into $b$ at the GDW in the unfolded picture.

In general, there exists certain $a\in \B$ such that $a$ cannot turn into any $b\in \B'$, {\it and} there exists certain $b'\in \B'$ such that $b'$ cannot turn into any $a'\in \B$, through the GDW. A special case, which we call {\it strict condensation wall}, is: For every $b \in \B'$, there exist ways (operators) at the GDW to turn $b$ into some $a\in \B$. That is, every anyon in $\B'$ can pass though a strict condensation wall and becomes  $a\in \B$. Note that $a$ may not be unique. In particular, let $A$ be a subset of anyons in $\B$, which are the anyons that can turn into $1\in \B'$. The set $A$ always contains $1\in \B$. On the other hand, not every anyon in $\B$ can pass through the GDW. In this sense,  strict condensation walls are asymmetric between $\B$ and $\B'$.

It was shown that strict condensation walls are completely classified by boson condensation\cite{Fuchs2002,Bais2009a,Kong2013,HungWan2015a}, in the following sense. The anyons in $A$ are those that are condensed in $\B$, and $\B'$ is the TO after condensing $A$ in $\B$. After condensation, all anyons in $A$ are mapped to $1\in \B'$, corresponding to the fact that anyons in $A$ can pass through the GDW and turn into $1\in \B'$. It is shown that if $A$ contains nontrivial anyons, $\B$ has more types of anyons and larger total quantum dimension than $\B'$, or mathematically, $\B'$ is a subcategory of $\B$.\cite{Fuchs2002,Bais2009a,Kong2013,HungWan2015a} The anyons in $\B$ that cannot cross the GDW correspond to those that will be confined in $\B'$. Since both $\B$ and $\B'$ are bTOs, the condensation must involve only bosons, in fact self-bosons, of $\B$. Note that the above is a over simplified story of boson condensation but this suffices for our purpose in this section. More detailed rules of boson condensation will be reviewed in Section \ref{subsubsec:revBC}.

One subtle case is that $\B=\B'$ and $A=\{1\}$. In this case, every anyon in $\B$ can move across the GDW and turn into some anyon in $\B'$, and every anyon in $\B'$ can pass through the GDW and turn into some anyon in $\B$ (i.e., the wall is symmetric). The trivial case is a uniform TO $\B=\B'$ everywhere. Nevertheless, there still exist nontrivial GDWs that permute the topological sectors in the topological order. As shown in Ref.\cite{Kitaev2012,Kong2013,HungWan2015a}, a nontrivial GDW between $\B$ and itself always implements a global symmetry on $\B$, in the sense that the wall acts as linear map of anyons of $\B$ that carry (not necessarily all) different representations of the global symmetry group. We thus dub such a GDW a \textit{symmetry wall}, which may be realized by a defect line in a lattice model or a real system\cite{Kong2013,Hung2013,HungWan2015a}. In the unfolded picture, a symmetry wall shall not be called a condensation wall, because the only ``condensed'' boson is $1$. In this paper, we always consider the case that $A$ contains anyons other than $1$. Hence, the subtlety of symmetry wall is not significant for us.

Now, back to the folded picture. One can see that a gapped boundary between $\B \bx \B'$ and the vacuum is also a strict condensation wall. Hence, through the folding trick, we see that general GDWs, not limited to strict condensation walls, can be understood as boson condensation in bTOs.

%
\subsection{Mutual Locality}\label{subsec:mutualLoc}
With the above briefing of boson condensation, we now discuss about for a given bTO $\B$, what anyons can condense and what are confined. The unconfined anyons form a new bTO $\B'$. The crucial notion condensability and confinement is mutual locality.

In Cooper pair condensation that leads to superconductivity, the condensed particles form the new vacuum of the system after the condensation. It is natural to expect in any kind of anyon condensation, the condensed anyons would become the new vacuum of the condensed phase. But what is a vacuum depends on the underlying degrees of freedom of the system under consideration. In the case with GDWs between bTOs, the underlying degrees of freedom are purely bosonic. As such, a well-defined vacuum of such a system would better be described by a bosonic theory, and in such a vacuum the only well-defined topological sector would have to be a trivial boson $1$. Having agreed on this, since the condensation describing a GDW between bTOs is to become a new purely bosonic vacuum that again contains only the trivial topological sector, the condensed anyons are expected to be mutual local with respect to the trivial boson $1$.  The question now is: What does it mean by being mutual-local with respect to $1$? Note that here we does not say \textquotedblleft with $1$" but instead \textquotedblleft with respect to $1$". The reason is that the trivial boson $1$ is indeed trivial and thus has trivial exchange statistics with anything else. That is, \textquotedblleft with respect to $1$" has a particular meaning, which can be understood if we know how one usually measures the topological spin of an anyon or braiding of two anyons.

To measure the braiding effect of two anyons, say $a$ and $b$, we create $a$ and its anti-anyon $\bar a$, and $b$ and $\bar b$ on a surface, then let $a$ and $b$ move on the surface such that they exchange their positions twice, and finally annihilate respectively $a$ with $\bar a$ and $b$ with $\bar b$. This procedure is equivalent to measuring the the interference between the state of the system after the braiding $a$ with $b$ and the state without braiding $a$ and $b$, which is captured by the $S$-matrix element $S_{ab}$. After appropriate normalization, what we actually measure is the monodromy matrix\cite{Bonderson2006,Bonderson2006a,Bonderson2007a,Bonderson2008,Nayak2008,Bais2009a}
\be\label{eq:monodromy}
M_{ab}=\frac{S_{ab}S_{11}}{S_{a1}S_{b1}}=\frac{\sum_c N^c_{ab}d_c\theta_c}{d_a\theta_a d_b\theta_b}.
\ee
Two anyons $a$ and $b$, not necessarily different, are \textit{mutually local}, i.e., winding one of them around the other has trivial effect on the state, if and only if $M_{ab}=1$. Keep in mind this is the definition in bTOs. Generic anyons $a$ and $b$ may have more than one fusion channels, and the monodromy matrix \eqref{eq:monodromy} is a weighed average of $M^c_{ab}$; hence, we can define a weaker notion of mutual locality: two anyons $a$ and $b$ are \textit{mutual-local via} $c$ if there exists a fusion channel $c\in a\x b$, such that
\be\label{eq:monoSingle}
M^c_{ab}=\frac{\theta_c}{\theta_a\theta_b}=1.
\ee
Two anyons mutual-local via certain fusion channel are also said to be partially mutual-local\cite{Bais2009a}.

We are now enabled to define that in a bTO, an anyon $a$ is mutual-local \textquotedblleft with respect to" the trivial boson $1$ if and only if $M^1_{a\bar a}=1$.
If $a$ is self-dual, then we need $M^1_{aa}=1$. Furthermore, if two distinct anyons $a$ and $b$, which are not anti-anyons of each other, can condense simultaneously, we would also require $a$ and $b$ being mutual-local via some anyon $c\in a\x b$, and $c$ better condense as well. This is reasonable because $a$ and $b$ will both become the new vacuum after condensation, and the vacuum is so trivial that $a$ and $b$ should better be at least partially mutual-local.

Clearly, the partially mutual-local condition $M^1_{a\bar a}=1$ implies that $\theta_a\theta_{\bar a}=\theta_a^2=1\Rightarrow \theta_a=\pm 1$. Namely, what can condense is either a self-fermion or a self-boson but nothing else if we demands such a definition of mutual-locality, with respect to a bosonic vacuum. Nevertheless, this condition is not strong enough for classifying GDWs between bTOs,  because if a self-fermion condensed, it would become a trivial fermion that exists only in an fTO. In other words, fermion condensation would result in  a fermionic vacuum rather than a purely bosonic one. Therefore, having been restricted to classifying GDWs between bTOs, one has to impose further constraints, such as, only self-bosons can condense. The mathematical formulation of such further constraints are reviewed in Section \ref{subsubsec:revBC} and not to be repeated here.

Now that the condition $M^1_{a\bar a}=1$ seems leaving us some room for condensing fermions, as argued above, to condense fermions in bTOs,  we need to consider the GDWs between a bTO and an fTO. Note that $M^1_{f\bar f}=1$ for any self-fermion, including the trivial fermion $1^f$ in any fTO. Moreover, $1^f$ and $1^f$ always fuse to $1$. Hence, the mutual-locality condition $M^1_{a\bar a}=1$ is well suited but at most suited for fermionic vacua. So, GDWs between bTOs and fTOs provide prefect systems for us to consider condensing self-fermions.

\subsection{Fermion Condensation and GDWs}\label{subsec:FandGDW}
Before we propose our ansatz for fermion condensation, let us provide a correspondence between fermion condensation and the GDWs between bTOs and fTOs, which is analogous to the correspondence between boson condensation and GDWs between bTOs.

\begin{figure}[h!]
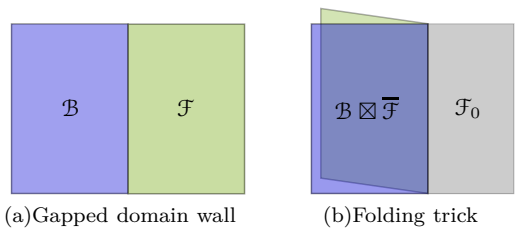

\subfigure[Gapped domain wall]{\label{subfig:gdw}
\gdw{\B}{\FF}
}
\subfigure[Folding trick]{\label{subfig:fold}
\gdwFold{\B\bx\overline{\FF}}{\FF_0}
}
\caption{(a) A bTO $\B$ and an fTO $\FF$ connected by a GDW via condensing some fermions in $\B$. Via the folding trick along the GDW, (a) is equivalent to (b) A topological order $B\bx\overline\FF$. sharing a gapped boundary with $\FF_0$, a fermionic vacuum.}
\label{fig:gdwAndFolding}
\end{figure}

We imagine condensing certain self-fermions in a bTO $\B$ and ending up with a topological order with a transparent (trivial) fermion besides a trivial boson, i.e., ending up with an fTO $\FF\supseteq\FF_0$. Although we do not know whether such mechanism of fermion condensation in certain order parameter space may exist or not, we do have a real space picture of the phase transition: the bTO $\B$ and fTO $\FF$ are connected by a GDW (Fig. \ref{subfig:gdw}), which belongs to some kind of strict condensation walls between bTOs and fTOs. The properties of such strict condensation walls actually ``define'' what we mean by fermion condensation and lead to our ansatz for fermion condensation, discussed in the next subsection. 

Discussed in Sec.~\ref{subsec:BCandGDW}, it is equivalent to look at the GDWs between two TOs via the folding trick.\cite{Kong2013,HungWan2014,HungWan2015a} As in the case of boson condensation in bTOs, one can fold the system in Fig. \ref{subfig:gdw} along the GDW between $\B$ and $\FF$. The result is the system in Fig. \ref{subfig:fold} as stacking a layer of $\B$ over a layer of $\overline\FF$, which is the time reversal of $\FF$. Since the trivial fTO $\FF_0=\{1,1^f\}\subseteq\FF$, condensing self-fermions $f'\in\B$ is equivalent to condensing the pairs $(f',1^f)\in \B\bx\FF$, which are effectively self-bosons.

The picture of equivalence delineated above not only helps understand fermion condensation in a bTO but also offers a physical realization of fermion condensation in a bTO. That is, in order to condense certain self-fermions $f'\in\B$, one can stack a layer of a trivial gapped fermion system $\FF_0$ with $\B$ to form $\B\bx\FF_0$ and then condense the effective self-bosons $(f',1^f)$ in this bilayer system (the trivial gapped fermion system would have chiral central charge $c=0$ such that stacking it with $\B$ does not change the total $c$). The result would be $\FF\bx_{\FF_0}\FF_0$. But according to the definition of stacking, $\FF\bx_{\FF_0}\FF_0\cong\FF$. One thus realizes the transition $\B\xrightarrow{\text{condense} f'}\FF$. We will see an example of this equivalence in Section \ref{subsec:su2}. 

One may however wonder given such an equivalence, one seems always able to work in the regime of boson condensation in the folded picture, which is true in principle. Nevertheless, we choose to start from scratch and understand fermion condensation in bTOs by itself. Conversely, the equivalent picture of boson condensation in fTOs can serve a double check of the principles of fermion condensation in bTOs to be found in the sequel.

\subsection{Ansatz of Fermion Condensation}

In this subsection, we write down our ansatz of fermion condensation, which can be generalized to more general anyon condensation in Appendix \ref{app:genAnyonCond}.
We take a formalism similar to that characterizing boson condensation, which is reviewed in Section \ref{subsubsec:revBC}.
\begin{ansatz*}
Fermion condensation in a bTO $\B$ is a composite object $A=n_a a\oplus n_b b\oplus\cdots\in \B$, where $a,b,\dots$ are anyons in $\B$ and $n_a$ the multiplicity (number of occurrences) of $a$ in $A$, that meet the following conditions.
\begin{itemize}
\item[1.] Unit: $\B$'s vacuum---trivial boson---$1\in A$, $n_1=1$.
\item[2.] Self-dual: For any $a\in A$, the anti-anyon $\bar a\in A$.
\item[3.] Self-locality: For any $a\in A$, $M^1_{a\bar a}=1$.
\item[4.] Closure: For any $a,b\in A$, there exists at least a $c\in a\x b$, such that $M^c_{ab}=1$ and $c\in A$.
%
\item[5.] Trivial associativity: $(A\x A)\x A=A\x(A\x A)$.
\end{itemize}
\end{ansatz*}
This ansatz is physically reasonable because the condensation would form the new fermionic vacuum of the system after the condensation, and a vacuum needs to have trivial statistics in order that the quasiparticle excitations have self and mutual statistics well defined with respect to the vacuum. A vacuum must also be closed under the fusion of its constituents. The trivial associativity condition is physically transparent but technically highly nontrivial. That an anyon may have to appear more than once in $A$ is also a consequence of trivial associativity, an example of which can be found in boson condensation in Ref.\cite{HungWan2014}. We shall look at this condition more closely upon necessity.

As discussed in Section \ref{subsec:mutualLoc} about mutual-locality, the self-locality condition in the ansatz forbids anything but self-bosons and self-fermions to condense. We repeat the argument here for clarity: $M^1_{a\bar a}=1\Rightarrow\theta_a\theta_{\bar a}=1$; however, since we know $\theta_a=\theta_{\bar a}$, we have $\theta_a^2=1\Rightarrow \theta_a=\pm 1$.

Anyon condensation in a two-layer system always fall into two scenarios: 1) the condensation is separate in each individual layer, and 2) some composite anyons made of those in each layer condense, such that the condensation can be viewed as occurring in a single-layer system. Anyon condensation in multi-layer systems follow likewise. Our Ansatz thus suffices in any bTOs. It is worth of mention that in classifying GDWs between bTOs via anyon condensation, the first scenario is usually neglected because it is rather trivial, as the two layers are completely uncoupled, independent systems.\footnote{In Ref.\cite{Lan2014}, the authos also include the trivial GDWs between uncoupled layers of two-layer systems.} As such, we shall ignore the first scenario too.

Although condensable anyons in a bTO can be either self-bosons or self-fermions, there exists anyon condensation that involves only self-bosons. Such cases are called (pure) boson condensation and have been thoroughly studied.\cite{Bais2002,Bais2009,Bais2009a,Hung2013,Kong2013,Gu2014a,HungWan2014,HungWan2015a}
In boson condensation, all condensates must be self-bosons and at least partially mutual-local with one another, which form what is known a twist-free commutative separable Frobenius algebra\cite{HungWan2015a} (CSFA), also referred to as a condensable algebra\cite{Kong2013}. For simplicity, we shall refer to the Frobenius algebras describing boson condensation bosonic Frobenius algebras, where \textquoteleft bosonic' encodes the extra conditions on top of Frobenius algebras.

Fermion condensation is more general than boson condensation. In fact, as we will show shortly, by our ansatz---in particular the trivial associativity condition---when a self-fermion $f$ condenses, if $f\x f$ may have to contain a self-boson $b$, such that $M^b_{ff}=1$ and $b$ condenses along with $f$. Thus, fermions that condense may be accompanied by bosons that have to condense simultaneously. In other words, boson condensation is a special case of fermion condensation. One thus can reasonably expect that the rules governing fermion condensation contain the rules of boson condensation as a subset, as to be seen. Since we always include the trivial boson in the set of condensed anyons for consistency, there does not exist pure fermion condensation.

%
\section{Fermion condensation in \MakeLowercase{b}TO\MakeLowercase{s}}

The main purpose of this section is to study properties of fermion condensation in bTOs, with the ansatz introduced in the previous section and with the GDW picture in mind. We establish a hierarchy principle for fermion condensation in bTOs. At the end of this section,  we also discuss about boson condensation and fermion condensation in fTOs.

\subsection{Principles of Fermion Condensation in bTOs}\label{subsec:pincipleFC}
Having had an expedition in  the physics and ansatz of fermion condensation, we are good to go back to the abstract picture Fig. \ref{subfig:gdw} to understand the rules and consequences of fermion condensation in bTOs.

Recall in the previous section we find that there can be pure boson condensation but fermion condensation in general involves boson condensation as well. Actually, we have a stronger relation between boson condensation and fermion condensation. First let us have a better understanding why boson condensation can always stand alone without invoking any fermions to condense. The reason roots in our fundamental ansatz. In order that all the conditions in the ansatz are satisfied, condensing certain anyons may force some other anyons to condense as well. Suppose two self-bosons (not necessarily different) $a$ and $b$ condense simultaneously, if $a\x b\ni c$ with $c$ a self-boson too, then the monodromy $M^c_{ab}\equiv 1$. So, $c$ potentially may condense; however, whether $c$ would be forced to condense or not further depends on the detail of the consequences of condensing $a$ and $b$, which is irrelevant for our discussion here. On the other hand, if $c$ is a self-fermion instead, then $M^c_{ab}\equiv -1\neq 1$ for sure, which is true from any self-bosons $a$ and $b$. Thus, condensing $a$ and $b$ cannot force $c$ to condense, and no matter what other self-bosons are fed to condense together with $a$ and $b$, $c$ would not be obliged to condense. Unless certain fermion $d$ that can condense along with $a$ and $b$ is mandated to condense by hand (i.e., by turning on a particular interaction to make $d$ condense), and $d$ can fuse with some condensed boson, say, $a$ for example, produces $c$. If so, because $M^c_{da}\equiv 1$, $c$ may be forced to condense now if its condensing does not violate other conditions in the ansatz. Therefore, if one would like to condense self-bosons only, one can leave the self-fermions alone without violating the ansatz.

Nevertheless, condensing self-fermions may force certain self-bosons to condense too. Suppose two self-fermions $a$ and $b$ (not necessarily different) condense, and there is a fusion channel $c\in a\x b$ that is a self-boson. Because $M^c_{ab}\equiv 1$ in this case, $c$ may be demanded to condense, which depends on further detail of condensing $a$ and $b$. We now show that in fermion condensation in a bTO, if there are two or more distinct self-fermions condense together, there must be concurrent self-boson condensation.

Consider a bTO $\B$ that supplies sufficient distinct self-fermions and self-bosons for our need. Let us condense only a single self-fermion $f\in\B$. Since the trivial boson $1$ is always included in any condensation, we denote the set of condensates temporarily by $A^F_0=1\oplus f$ and see if it fulfills the ansatz. Immediately, we see that $f$ has to be self-dual, i.e., $f\x f\ni 1$; otherwise, by the ansatz $\bar f\neq f$ must also condense. Now that $f$ is self dual, we have at least $M^1_{ff}\equiv 1$, and thus $f$ can indeed condense. All conditions in the ansatz are satisfied. Hence, $A^F_0$ is a well-defined fermion condensation; in fact it is the smallest possible fermion condensation, and the only fermion condensation that does not contain any nontrivial self-boson. We dub $A^F_0$ \textit{minimal fermion condensation}. We will justify the meaning of this minimality further in Lemma \ref{lem:minFC}. An example is $A^F_0=1\oplus\psi$ in the Ising topological order $\B_\text{Ising}=\{1,\sigma,\psi\}$, where $\psi$ is the only self-fermion.

Before we build larger anyon condensation consisting of more distinct anyons, let us gain better understanding of the trivial associativity condition $(A\x A)\x A=A\x(A\x A)$ in the ansatz. Recall that condensation $A$ is an object of the UMTC $\B$ describing a bTO, namely, it is a composite anyon. If $A$ condenses, it would become the vacuum of the the new topological order after the condensation. In contrast to nontrivial anyons---nonlocal objects---in a topological order, the vacuum sector is local. In the standard picture of topological orders, nontrivial anyons are created by nonlocal string operators and thus are sitting at the ends of strings. The vacuum, however, is associated with local operators and hence not attached to any string or equivalently speaking is attached to a trivial string. Fusion between nontrivial anyons can be viewed as interaction between the strings the anyons are attached to. Fusion is associative, i.e., for three anyons $a,b$, and $c$, $a\x(b\x c)\cong (a\x b)\x c$. The isomorphism $\cong$ indicates that its two sides, $a\x(b\x c)$ and $(a\x b)\x c$ are two Hilbert spaces in general. But isomorphism is not equality, as reconnecting the strings involved in the fusion interaction to go from $a\x(b\x c)$ to $(a\x b)\x c$ changes the basis of the isomorphic Hilbert spaces. Such a basis change is a nontrivial linear transformation.

Equipped with this understanding, the trivial associativity in our ansatz simply states that for condensation $A$, to become the new vacuum, the linear transformation from $(A\x A)\x A$ to $A\x(A\x A)$ is trivial---an identity. It is however highly nontrivial to prove that an arbitrarily constructed $A$, which clearly satisfies all but not obviously the trivial associativity condition, actually does satisfy the trivial associativity. For this, in general one would need the full topological data of the bTO where $A$ belongs to. Nevertheless, fortunately, if an $A$ does not meet the trivial associativity, it is rather simple to show. In fact, as proven in the Appendix \ref{app:trivialAssoc}, trivial associativity of $A$ leads to a necessary criterion for judging whether $A$ may be well-defined condensation. For clarity, we repeat this necessary condition here.
\begin{lemma}\label{lem:associativity}
For any nontrivial anyon $a\in A$, $A$ being well-defined anyon condensation, if $a$ is a nonsimple current, i.e., $d_a>1$, then there exists at least a nontrivial self-boson $b\in a\x\bar a$, such that $b\in A$.
\end{lemma}

Note that in the lemma, $b$ is not necessarily different from $a$ or $\bar a$. Also note that the anyons $a$ under consideration are either self-bosons or self-fermions. 
The nontrivial anyon $b$ must be a self-boson no matter what $a$ is because it is the only way of guaranteeing $M^b_{a\bar a}=1$. Lemma \ref{lem:associativity} leads to an important result. Let us look at the minimal fermion condensation $A^F_0=1\oplus f$ we brought up earlier, where we only require that $f\x f\ni 1$; however, Lemma \ref{lem:associativity} readily demands that $d_f=1$ and hence $f\x f=1$, as otherwise, $A^F_0$ must contain at least another self-boson for it to be well-defined condensation. We thus have the following lemma.
\begin{lemma}\label{lem:minFC}
The fermion $f$ in  minimal fermion condensation $A^F_0=1\oplus f$ must be a simple current, i.e., it has quantum dimension $d_f=1$, and therefore satisfies the fusion rule $f\times f=1$. 
\end{lemma}

Note that the self-fermion $f$ in $A^F_0$ is not a universal particle that is the same in all bTOs; rather, it is a particular self-fermion in any given bTO $\B$, which has certain braiding with other anyons in $\B$. Given a different bTO $\B'$, $f$ is certainly some particular self-fermion of $\B'$ and has nothing to do with that in $\B$. This $A^F_0$ will play a key role in fermion condensation. To that end, let us add more condensable anyons to build larger fermion condensation. There are the following three ways of doing this enlargement, a consistent discussion of which requires us not to assume $f$ being a simple current.

We again take a single self-fermion $f$. First, we can assume that $f$ is not self-dual. So, there exists an $\bar f\neq f$ in $\B$. The ansatz requires $\bar f$ to condense with $f$. But this causes an issue: since $f$ is not self-dual, $f\x f\not\ni 1$; hence, we need another channel $b\in f\x f$ with $M^b_{ff}=\theta_b/\theta_f^2=1$, such that $f$ meets the ansatz and can continue to condense. It is clear that $\theta_b=1$, i.e., $b$ is a nontrivial self-boson. As such, by the ansatz our condensation becomes $A^F_1=1\oplus f\oplus b\oplus\bar f$. Such that this $A^F$ is closed under fusion, we need the fusion rules $f\x f\ni b$, $f\x b\ni\bar f$, $b\x b\ni 1$, $\bar f\x b\ni f$, $\bar f\x\bar f\ni b$, and $f\x\bar f\ni 1$. The anyons that are neglected in these fusion products are irrelevant because they do not have to condense to make $A^F_1$ well defined. We demand $b$ self-dual also to avoid introducing more anyons to condense. But we refrain from discussing whether these condensable anyons are simple currents or not, which is case dependent. Examples of $A^F_1$ can be found in Abelian bTOs.

Second, we can still let $f$ being self-dual but add extra anyons to enlarge the condensation. For our purposes, let us add exactly one more self-fermion $f'\in\B$ to condense. This implies that $f'$ is self-dual too. Again, an issue would arise if we only condensed $f$ and $f'$. Since $f\neq f'$ and they are both self-dual, $f\x f'\not\ni 1$. We thus need at least another channel $b\in f\x f'$ with $M^b_{ff'}=1$ for $f$ and $f'$ can simultaneously condense, which forces $b$ being a self-boson. Moreover, $A^F$ must be closed under fusion. So, either $b$ or some other self-boson similar to $b$ must condense. To limit the size of the condensation, we simply make $b$ condense and self-dual. We then have ${A^F_1}'=1\oplus f\oplus b\oplus f'$ with each element self-dual and extra fusion rules $f\x b\ni f'$, $f\x f'\ni b$, and $f'\x b\ni f$. An example is ${A^F_1}'=1\oplus \psi 1\oplus\psi\bar\psi\oplus 1\bar\psi$, where $b=\psi\bar\psi$, in the doubled Ising topological order $\B_{\rm{Ising}}\bx\B_{\overline{\rm{Ising}}}$.
The stacking in this example is nontrivial because condensing $\psi\bar\psi$ requires coupling the $\psi$ and $\bar\psi$ respectively of the two-layers. That is, $\B_{\rm{Ising}}\bx\B_{\overline{\rm{Ising}}}$ is treated as a single-layer system. Condensing ${A^F_1}'$ will appear to be rather simple as soon as we complete developing the rules of fermion condensation.

Third, instead of adding more self-fermions to condense along with the $f$, we can try to add only one more self-boson, say, $b\in\B$ to the condensation. We need $f$ and $b$ being both self-dual so as not to involve any more anyons to condense. It is straightforward to check that the fermion condensation defined by $A^F_2=1\oplus b\oplus f\in\B$ satisfies the ansatz if $b$ and $f$, apart from being self-dual, are also subject to the fusion rules $f\x b\ni f$, $f\x f\ni b$, $b\x b\ni b$, and $f\x b\ni b$. We will show an example of $A^F_2$ condensation in Section \ref{subsec:DD3}.

Without further enlarging $A^F$, the above discourse has already shown clearly that any $A^F$ larger than $A^F_0$ must contain nontrivial self-bosons. Such self-bosons can surely condense simultaneously, and as argued earlier, their condensing does not force the self-fermions in the same $A^F$ to condense although the converse is true. These self-bosons are included in the condensation to make fermion condensation well-defined. In other words, the self-bosons in an $A^F$ form legal boson condensation on their own. According to Ref.\cite{HungWan2015a}, boson condensation in a bTO is specified by a bosonic Frobenius algebra $A^B$ whose elements are simultaneously condensable self-bosons, including the trivial boson $1$. Although we lack a completely rigorous mathematical account of the algebraic structure of $A^F$, our ansatz and the discussion above conjectures that $A^F$ would still be a Frobenius algebra but not a bosonic one because the twist-free condition and commutativity condition seem relaxed, as discussed in Appendix \ref{app:Frobenius}. Let us then from now on call $A^F$ a \textit{fermionic Frobenius algebra}.  The long exposition above readily establishes the following lemma.
\begin{lemma}\label{lem:superFA}
Any fermionic Frobenius algebra $A^F$ characterizing fermion condensation in a bTO contains a bosonic Frobenius algebra $A^B\subset A^F$ as a subalgebra. Importantly, the self-fermions in an $A^F$ cannot form a closed algebra under fusion on their own.
\end{lemma}

The trivial case is that $A^B_0=1\subset A^F_0$. Note that an $A^B$ is not only a Frobenius algebra but a twist-free CSFA. Twist-free requires $A^B$ contain only self-bosons. Commutativity is the usual one for algebra multiplication. Separability admits a direct sum presentation of $A^B$ in the bTO $\B$, which is a UMTC. That is, $A^B=1\oplus_b b$ as a single (but not simple) object in $\B$. We take these conditions for granted and simply call an $A^B$ a Frobenius algebra to avoid clutter. We also remark that although the fermions in an $A^F$ do not form a closed algebra, sometimes, certain simple current fermions together with the trivial boson of $A^F$may form a condensable subalgebra of $A^F$. We shall see an example of this in Section \ref{subsec:su2}.    

Lemma \ref{lem:superFA} has a crucial consequence; it leads to our first and main principle as follows of doing fermion condensation in bTOs.

\textbf{Hierarchy Principle}. To do fermion condensation specified by a fermionic Frobenius algebra $A^F\supseteq A^B$ in a bTO $\B$, one can always first perform the boson condensation $A^B$ entirely, applying the rules of boson condensation, and then condense the necessary self-fermions.

This principle needs more elaboration, and interesting results will follow. First, if $A^F=A^F_0$, there is no nontrivial boson condensation involved, and one shall directly proceed to condense the only self-fermion in $A^F_0$. We will get to the rules of such condensation later.

Second, if $A^F\neq A^F_0$, $A^B$ is not trivial. One can then proceed to condense $A^B$ first. What is next? The answer is surprising and simple: Having condensed $A^B\subset A^F$, what remains to condense is merely an $A^F_0$, i.e. the previously defined minimal fermion condensation, which contains only a simple current self-fermion.
Certainly, condensing $A^B\subset A^F$ would alter the anyon content of the original bTO $\B$ and hence the self-fermions in the $A^F$ but we need not to know any such detail to proceed. According to the theory of boson condensation\cite{Bais2009a,HungWan2015a,Kirillov2002,Fuchs2002}, we know that condensing $A^B$ turns $\B$ into a smaller bTO $\B'$, the details of which is irrelevant for now. As such, the $A^F\supset A^B$ would become some ${A^F}'$ as a composite object in $\B'$. Our claim is that ${A^F}'\equiv A^F_0$. This can be proven by contradiction as follows.

There are three steps. 1) The object ${A^F}'$ must contain at least one self-fermion. Suppose not, i.e., ${A^F}'=1$, since boson condensation does not induce any fermion condensation, this supposition would imply that there were not any self-fermion involved in the $A^F$ anyway in the first place, which is beyond our consideration. 2) There is exactly one self-fermion in ${A^F}'$. If not, according to our discussion before, there has to be certain self-bosons remaining in ${A^F}'$ in order that ${A^F}'$ is well-defined, which is contradictory to that we have already condensed all the self-bosons in $A^F$. 3) By Lemma \ref{lem:minFC}, fermion condensation of exactly one self-fermion must be $A^F_0$, namely, ${A^F}'\equiv A^F_0$. These establish the first half of the following theorem.
\begin{theorem}\label{theo:FCdecomp}
Any fermion condensation $A^F$ in a bTO $\B$ always admits the decomposition \be\label{eq:FCdecomp}
A^F=A^B\orbx A^F_0,
\ee
where $A^B$ is the bosonic part of $A^F$. Such condensation turns $\B$ into an fTO $\FF$, such that the dimensions of $\B$, $\FF$, and $A^F$ satisfy the relation
\be\label{eq:dimRel}
D_\FF=\sqrt{2}\frac{D_\B}{\dim A^F}=\frac{D_\B}{\sqrt{2}\dim A^B}.
\ee
\end{theorem}

In this theorem, particularly in Eq. \eqref{eq:FCdecomp}, we use the notation $\orbx$, which is similar to a stacking operation in only one direction. We use this notation because we still have not fully understood the rigorous algebraic structure of $A^F$ and the representation category of $A^F$, which we call a fermionic Frobenius algebra. Fortunately we find that the effect of $A^F$ condensation in $\B$ is equivalent to condensing a twist-free CSFA $A^B\subset A^F$ followed by minimal fermion condensation $A^F_0$, also a minimal fermionic Frobenius algebra. Condensing $A^B$ first results in an intermediate bTO $\B'$, the simple current self-fermion $f$ in the $A^F_0$ is an anyon of $\B'$ rather than an anyon of $\B$. The relation between this $f$ and the self-fermions in $A^F$ is nontrivial and depends on the details of $\B$ and $A^F$, which we shall address this in an example later. We simply denote such a two-step procedure of condensing $A^F$ by Eq. \eqref{eq:FCdecomp}, as stacking $A^F_0$ on top of $A^B$. We also use $\orbx$ to emphasize that $A^F_0$ is not a fermionic Frobenius algebra in $\B$ but in $\B'$; otherwise, we would have simply used the usual tensor product $\otimes$.

Conversely, any hierarchical fermion condensation $A^B\orbx A^F_0$ can be lifted to a nontrivial fermion condensation $A^F$. We will com back and elaborate on this scenario in Section \ref{subsec:otherGDW}.

We now prove the second half of Theorem \ref{theo:FCdecomp}, namely Eq. \eqref{eq:dimRel}, which is the relation between the total quantum dimensions of the original bTO $\B$, that of the fTO $\F$ after the condensation, and the dimension of the condensation $A^F$. For a boson condensation $A^B$, which is a Frobenius algebra, its dimension has a canonical definition $\dim A^B=\sum_{a\in A^B}d_a$.\cite{HungWan2015a,Kirillov2002,Fuchs2002} Here, since $A^F$ is still a Frobenius algebra, we have $\dim A^F=\sum_{a\in A^F}d_a$. Then, we have $\dim A^F_0=2$. Since the decomposition $A^F=A^B\orbx A^F_0$ is indeed analogous to stacking, we naturally have $\dim A^F=\dim A^B \dim A^F_0=2\dim A^B$. According to the theory of boson condensation\cite{HungWan2015a}, the dimensions of $\B$ and $\B'$, due to boson condensation $A^B$, are related by $D_{\B'}=D_\B/\dim A^B$. Hence, what remains to show is that
\be\label{eq:dimRelPrime}
D_\FF=\sqrt{2}\frac{D_{B'}}{\dim A^F_0}=\frac{D_{\B'}}{\sqrt{2}}.
\ee
We defer proving this until a little later when we fully understand $A^F_0$ condensation. Before that, let us have a quick review of the rules of boson condensation.

\subsection{Review of Boson Condensation}\label{subsubsec:revBC}
For a full account of boson condensation, please see for example Ref.\cite{Bais2009a,HungWan2015a} for details. A boson condensation in a parent bTO $\B$ is described by a bosonic Frobenius algebra, i.e., a twist-free CSFA $A^B$, which is a composite object in $\B$. This $A^B$ fulfills all the conditions in our ansatz, and on top of this, each constituent of $A^B$ must be a self-boson, which is the twist-free condition. Besides, $A^B$ is commutative (see appendix). Having condensed $A^B$, the entire algebra $A^B$ is projected to the new trivial boson---new vacuum. The intermediate result is a tensor category $\T$. The simple objects of $\T$ are defined with respect to the new vacuum $1_\T\mapsfrom A^B$ and are generally composite objects of the original $\B$.

So, the next step is to find out all the simple objects of $\T$. To do so, note that there exists a functor map $F:\B\rightarrow\T$, such that for any anyon $a\in\B$, $F(a)\xeq{\rm{def}}a\x A^B\in\T$. Nevertheless, not every simple object of $\T$ is mapped from an anyon of $\B$ by the functor $F$. It turns out that for any $a\in\B$, if $d_a<2$, $F(a)\in\T$ is a simple object. Using this, one can then list out $F(a)$, $\forall a\in\B$ and search for the simple objects of $\T$ recursively.

If a simple object $t\in\T$ takes the form $t=F(a)$ for some $a\in\B$, then $d_t=d_a$; if not, say, for instance $t=a\oplus b\neq F(c)$, $\forall c\in\B$, then $d_t=(d_a+d_b)/\dim A^B$. Keep in mind that $d_t$ is the quantum dimension of $t$ defined in $\T$. The phenomenon that $t\in \T$ is a direct sum of certain anyons of $\B$ is called \textit{identification}, in the sense that the anyons appearing in the direct sum are identified as the same simple object in $\T$.

If an anyon $a\in\B$ appears in more than one simple objects of $\T$, we say that $a$ \textit{splits}
\be\label{eq:split}
a\rightarrow\sum_{i=1}^m n^i_a a_i,
\ee
where $n_a^i\in\Z_{+}$ is the multiplicity of species $a_i$ that are simple objects of $\T$. Practically, $a$ splits if $a$ is a \textit{fixed point} of its fusion with any nontrivial condensed self-boson, i.e., if $a\x c\ni a$ for any $c\in A^B$. Splitting preserves quantum dimension:
\be\label{eq:splitQDconserve}
d_a=\sum_{i=1}^m n^i_ad_{a_i}.
\ee
Besides, splitting and fusion commute:
\be\label{eq:splitFusionCommute}
\biggr(\sum_i n^i_a a_i\biggr)\x \biggr(\sum_j n^j_b b_j\biggr) =
\sum_{c,k}N^c_{ab} n^k_c c_k,
\ee
which directly follows from that $F$ is a tensor functor. Clearly, identification and splitting generally can hardly be two separated, independent phenomena.

Nevertheless, $\T$ is not the final result of $A^B$ condensation. Since $1_\T\mapsfrom A^B$ is the new vacuum, a well-defined new anyon has to be mutual local with respect to $1_\T$. Among all simple objects of $\T$, only those coming from the anyons of $\B$ that are mutual local with the algebra $A^B$ may be mutual local with $1_\T$, and the rest simple objects of $\T$ are said to be confined. In general, because of the complexity of $A^B$ and the splitting of anyons, it is not easy to apply this criterion of confinement. A convenient trick is offered by Bais \textit{et al}\cite{Bais2009a}: Consider $t\in \T$ and $t=a\oplus b\oplus\dots$, where $a,b,\dots\in\B$, the $t$ is confined if any two constituents in this combination have topological spins differ by a noninteger, say, for example, if $h_a-h_b\not\in\Z$. This is physically reasonable because a new anyon must have well-defined topological spin. Topological spins that differ by merely an integer are regarded the same, as they differ by a true boson.

Given a UTMC $\B$, after condensing $A^B\in\B$, all unconfined anyons form a child UTMC and hence a bTO $\B'$. The relation between various dimensions is $D_{\B'}=D_\B/\dim A^B$, as aforementioned. One sees that this relation between dimensions is merely a special case of our formula \eqref{eq:dimRel}. The confined sectors do not just disappear but become defects in $\B'$ and in many cases generate a global symmetry acting on $\B'$. Hence, the child topological order is in fact a symmetry-enriched topological order\cite{Hung2013,Gu2014a}. Typically, a global symmetry on $\B'$ arises when the condensed self-bosons are all simple currents because simple currents form a group under fusion, so do the confined sectors due to the condensation. Condensation of nonsimple currents may also yield a global symmetry on $B'$ but this is not guaranteed because the confined sectors due to nonsimple current condensation do not form a group under fusion in general. The confined sectors can also be viewed as symmetry fluxes in $\B'$. Gauging the global symmetry on $\B'$ takes $B'$ back to its parent bTO $\B$.

A typical example is $\B= \B_{\rm{Ising}}\bx\B_{\overline{\rm{Ising}}}=\{1,\sigma,\psi\}\bx\{1,\bar\sigma,\bar\psi\}$ and $A^B=1\oplus\psi\bar\psi\in\B$. After condensing this $A^B$, we have $\B'=\{1,e,m,\epsilon\}$, which is the $\Z_2$ toric code bTO. Here, the splitting occurs as $\sigma\bar\sigma\rightarrow e+m$, and $\psi 1$ and $1\bar\psi$ are identified to be $\epsilon$. The global symmetry generated on the $\Z_2$ toric code is a $\Z_2$ symmetry whose action exchanges $e$ and $m$. Gauging this $\Z_2$ symmetry takes the toric code back to the doubled Ising.
\subsection{Minimal Fermion Condensation}
Having recalled the protocol of boson condensation, we are now good to study minimal fermion condensation. 
Consider a generic bTO $\B$ that contains a minimal fermionic Frobenius algebra $A^F_0=1\oplus f\in\B$. We emphasize that this $f$ has the properties $f\x f=1$, $d_f=1$, and $\theta_f=-1$. Like boson condensation, after condensing $A^F_0$, $A^F_0$ would be projected to the new vacuum. The only and key difference here is that in an fTO $\FF$, the vacuum contains two trivial sectors, a trivial boson and a trivial fermion, which are both mutual local with any other anyons in the fTO. In other words, there is a trivial fTO $\FF_0=\{1,1^f\}$, as defined earlier, embedded in $\FF$. In general, such an embedding is not trivial. In other words, generally $\FF$ does not admit a stacking decomposition $\FF=\FF_0\bx \B'$ for any bTO $\B'$. This $\FF_0$ has total quantum dimension $D_{\FF_0}=\sqrt{2}$ and a canonical modular $S$-matrix, which is however degenerate,
\be
S_{\FF_0}=\frac{1}{\sqrt{2}}
\bpm
1 & 1\\
1 & 1
\epm.
\ee
Hence, $\FF_0$ is not a modular tensor category.

Condensing $A^F_0\in\B$ means to identify $A^F_0$ with $\FF_0$ in the resultant fTO $\FF$ after the condensation. All nontrivial anyons in $\FF$ must be defined with respect to $\FF_0$. Similar to boson condensation, identification and splitting of $\B$ anyons can happen too in fermion condensation, so do confinement and unconfinement.

Let us first focus on confinement. For $A^F_0$ condensation, the scenarios are much simpler. Because $f\x f=1$, it can neither split or be identified with anything else in $\B$. That is, $f$ must unanimously become the $1^f$ in $\FF$. Moreover, since $f$ is a simple current, the fusion between $f$ and any anyon of $\B$ produces exactly one anyon of $\B$. Hence, to see whether a $\B$ anyon $a$ will be confined due to $f$ condensation, it suffices to check the mutual monodromy $M_{af}=M^b_{af}$, where $b$ is the sole fusion channel $b=a\x f$. By Eq. \eqref{eq:monodromy}, we have $M_{af}=S_{af}/S_{1a}=D_\B S_{af}/d_a$. Hence, the monodromy $M_{af}$ is completely determined by the $S$-matrix element $S_{af}$. Clearly, $M_{af}=1$ only if $S_{af}=d_a/D_\B$. Before we make this claim formally, however, let us check what $S_{af}$ can be most generally. For this, we shall deploy the Verlinde formula\cite{Drinfeld2010}\footnote{These two versions of Verlinde formula are equivalent for UMTCs because the $S$-matrices in such cases are unitary and nondegenerate.}
\begin{align}
S_{ab}S_{ac} &=d_a\sum_{e\in\B}N^e_{bc}S_{ae},\label{eq:verlinde1}\\
N^c_{ab}&=\sum_{e\in\B}\frac{S_{ae}S_{be}S_{\bar c e}}{S_{1e}},\label{eq:verlinde2}
\end{align}
where $a,b,c,e\in\B$. Setting in Verlinde formula \eqref{eq:verlinde1} $b=c=f$, the condensed self-fermion, which enforces $e=f\x f=1$, we have
\be
(S_{af})^2=(S_{a1})^2=(\frac{d_a}{D_\B})^2\Longrightarrow S_{af}=\pm \frac{d_a}{D_\B}.
\ee
That is, all anyons in $\B$ are graded by the simple current fermion $f\in\B$ into two disjoint sets of anyons,
\be\label{eq:grading}
\begin{aligned}
&U_\B=\{1,f,u_1,u_2,\dots|u_i\in\B, S_{u_i f}=d_{u_i}/D_\B\},\\
&C_\B=\{c_1,c_2,\dots|c_i\in\B, S_{c_i f}=-d_{c_i}/D_\B\}.
\end{aligned}
\ee
Now clearly, the set $C_\B$ will be confined, whereas set $U_\B$ will be unconfined, after $A^F_0$ condensation. In the picture of GDWs, the confined anyons are those cannot penetrate the GDW between $\B$ and $\FF$, while the unconfined ones can cross the GDW from $\B$ and become (possibly after recombination) well-defined anyons in $\FF$.

An important result follows from the above grading. In Verlinde formula \eqref{eq:verlinde2}, let $a=c=1$ and $b=f$, and since $1\x f=f$, we have
\be\nonumber
\begin{aligned}
0=N^1_{1f}&=\sum_{e\in\B}S_{1e}S_{fe}\\
&=\sum_{e\in U_\B}S_{1e}S_{fe}+\sum_{e\in C_\B}S_{1e}S_{fe}\\
&=\sum_{e\in U_\B}\frac{d_e^2}{D_\B^2}-\sum_{e\in C_\B}\frac{d_e^2}{D_\B^2},
\end{aligned}
\ee
where use of the grading \eqref{eq:grading} is made. Thus we obtain
\be\label{eq:qdGrading}
\sum_{e\in U_\B}d_e^2=\sum_{e\in C_\B}d_e^2.
\ee
Since the anyons in the set $U_\B$ will become (possibly after recombination) anyons in the fTO $\FF$ after $A^F_0$ condensation in $\B$, we obtain
\be\label{eq:dimRelAF0}
D_\FF=\frac{D_{\B}}{\sqrt{2}}=\sqrt{2}\frac{D_{B}}{\dim A^F_0}.
\ee
Since $\B$ is an arbitrary bTO that supplies an $A^F_0$, we conclude that Eq. \eqref{eq:dimRelPrime} is correct\footnote{We note that a similar proof of this relation can be found in Ref.\cite{Lan2015} but the context there was the modularization of a fTO.}, and so is Eq. \eqref{eq:dimRel}. This completes the proof of Theorem \ref{theo:FCdecomp}.

Now let us study splitting and identification of anyons through $A^F_0$ condensation in $\B$. In boson condensation, splitting of an anyon may occur only when the anyon is a fixed point of its fusion with any nontrivial condensed self-boson. In fermion condensation, we expect the similar phenomenon, and in particular for $A^F_0$ condensation, we can address splitting straightforwardly. Suppose $a\in\B$ satisfies $a\x f=a$, i.e., $a$ is a fixed point. The definition of $S$-matrix \eqref{eq:S} implies that
\be\nonumber
S_{af}=\frac{\theta_a d_a}{\theta_a\theta_f D_\B}\equiv -\frac{d_a}{D_\B}.
\ee
Thus, $a\in C_\B$ and is confined no matter it splits or not.

As to the phenomenon of identification, consider two distinct anyons $a,b\in\B$, such that $a\x f=b$. Again the $S$-matrix \eqref{eq:S} implies that
\be\nonumber
\begin{aligned}
&S_{af}=\frac{\theta_a d_a}{\theta_b\theta_f D_\B}=-\frac{\theta_a d_a}{\theta_b D_\B}\\
\Longrightarrow\ &S_{af}=\frac{d_a}{D_\B}\ \text{iff}\ |h_a-h_b|\in\Z_+/2.
\end{aligned}
\ee
Because $a\x f=b$ also implies $b\x f=a$, as $f\x f=1$, an anyon may only be identified with exactly one other anyon. The derivation above shows that any two identified anyons whose topological spins differ by a half integer will be unconfined and become anyons in the final fTO $\FF$. Such identification in fermion condensation has a subtlety.

Recall that in boson condensation, if two anyons are identified and unconfined, their topological spins differ by merely an integer, or in other words, they differ by a trivial boson that is identified with the condensed self-boson. If we are not concerned with the global symmetry generated on the unconfined anyons by the boson condensation,\footnote{Strictly speaking, boson condensation in a bTO results in not a pure topological order but a symmetry-enriched topological order\cite{Gu2014a,HungWan2015a}. Anyons identified due to boson condensation may still carry different global symmetry representations.} the anyons that are identified are truly the same anyon in the resultant topological order.

Back in the fermion condensation under consideration, the new vacuum after condensing $A^F_0$ is not anything structureless but consists of two superselection sectors that differ by fermion parity. The vacuum of a physical fermionic system respects locality would be in either but not both of the two superselection sectors, such that fermion parity is a good quantum number. Hence, two anyon excitations that differ by fermion parity, i.e. that their topological spins differ by a half integer, can be well distinguished by local experiments and thus would not be truly identified. Rather, in $A^F_0$ condensation, two anyons $a$ and $b$ related by $a\x f=b$ may be considered as a doublet in which they have opposite fermion parity. In other words, there is a $\Z_2^f$---fermion parity symmetry---generated on $\FF$ due to $A^F_0$ condensation. Actually, in any fTO, the anyons are grouped in \textquotedblleft super pairs", and in each super pair the two anyons are \textquotedblleft super partners" of each other.\footnote{This is merely an analogy but does not necessarily imply supersymmetry.}

For non-prime fTOs, the Hilbert space of a super pair admits a simple decomposition into subspaces of even and odd fermion parities respectively. Consider a non-prime fTO $\FF_{\rm{np}}=\B_{\rm{p}}\bx\FF_0$ for some bTO $\B_{\rm{p}}$. Then the anyon content of $\FF_{\rm{np}}$ takes the form $\{(a,a^f)|a\in\B_{\rm{p}}\}$. Since $\B_{\rm{p}}$ is a well-defined bTO, the fusion of any two anyons $a,b\in\B_{\rm{p}}$ never produces neither of their super partners, and vice versa. The fusion of a $\B_{\rm{p}}$ anyon with its super partner always yields the super partners of $\B_{\rm{p}}$ anyons. This manifests the decomposition of the Hilbert space of a super pair. In contrast, in a primitive fTO, the Hilbert space of a generic super pair, say, $(b,b^f)$ is rather nontrivial and admits no decomposition into fermion parity even and odd subspaces. This is because the fusion $b\x b$ in general can produce both anyons and their super partners as well. So do the fusion $b^f\x b^f$ and $b\x b^f$.

Now that any fermion condensation $A^F$ can be decomposed into a boson condensation $A^B$ followed by solely $A^F_0$ condensation, the $\Z_2^f$ symmetry is guaranteed to exist after the condensation. The boson condensation $A^B$ may also generate a global symmetry $G_s$ on $\FF$. If this happens, the total symmetry would be certain $\Z_2^f$-extension of $ G_s$. A trivial extension would be $\Z_2^f\x G_s$.

The last aspect of $A^F_0$ condensation in $\B$ regards the fusion rules in the child fTO $\FF$. By the commutativity between splitting/identification and fusion, the fusion between unconfined, namely anyons in $\FF$ cannot yield any confined anyons, also as the latter are excluded from the spectrum of $\FF$ in the first place. Since there is no real identification of $\B$ anyons through $A^F_0$ condensation, all unconfined $\B$ anyons remain as individual anyons in $\FF$ despite being grouped into super pairs. Therefore, the fusion rules of the anyons in $\FF$ are simply the same as those of these anyons in $\B$ except that the $f\in A^F_0$ that appear in $\B$'s fusion rules must now be replaced by $1^f$ in the corresponding fusion rules of $\FF$.

For clarity, we can summarize the above  discussion in the following five (not mutually exclusive) rules of $A^F_0$ condensation in $\B$.
\begin{enumerate}
\item Any $a\in\B$ with $S_{af}=d_a/D_\B$ is unconfined and otherwise confined.
\item Any $a\in\B$ with $a\x f=a$ is confined.
\item Any $a,b\in\B$ with $a\x f=b$ are unconfined if $|h_a-h_b|\in\Z_+/2$ and otherwise confined. A pair of such $a$ and $b$ becomes super pairs in $\FF$. 
\item The fTO $\FF$ due to $A^F_0$ condensation in $\B$ contain all the unconfined sectors. Dimensions of $\FF$ and $\B$ satisfy Eq. \eqref{eq:dimRelAF0}.
\item Fusion rules in $\FF$ are the same as those in $\B$ with $f$ replaced by $1^f$.
\end{enumerate}
\subsection{Other GDWs between fTOs}\label{subsec:otherGDW}
We have studied fermion condensation in a bTO $\B$ that takes $\B$ to an fTO $\FF$. Such a picture is equivalent to a GDW between $\B$ and $\FF$ on which the condensed anyons become excitations that can be created and annihilated by local operators. The confined anyons in the condensation picture are those $\B$ anyons that cannot go from $\B$ through the GDW into the fTO $\FF$, while the unconfined one are those who can move between $\B$ and $\FF$ through the GDW, which acts as a transformation on the unconfined anyons. Bearing this and the discussion so far in mind, we now elaborate on an important remark on Theorem \ref{theo:FCdecomp} we made in Section \ref{subsec:pincipleFC}.

Suppose a bTO $\B$ admits a boson condensation $A^B$ and breaks into a child bTO $\B'$. Assume that this $\B'$ happens to bear a minimal fermion condensation $A^F_0$ and can break into an fTO $\FF$. This guarantee that $\B$ admits a nontrivial $A^F$ that directly breaks it into $\FF$. In other words, there exists an $A^F\in\B$ that can be decomposed as $A^B\orbx A^F_0$. We shall see an example of this in Section \ref{subsec:su2}. The following figure depicts this scenario from the perspective of GDWs.
\begin{figure}[h!]
\subfigure[]{\scalebox{0.9}{
\gdwTwo{\B}{\B'}{\FF}{W_{\B\B'}}{W_{\B'\FF}}}
}
\subfigure[]{\scalebox{0.9}{
\gdwNo{\B}{\FF}{W_{\B\FF}=W_{\B\B'}W_{\B'\FF}}}
}
\caption{(a) A GDW between $\B$ and $\B'$ and one between $\B'$ and $\FF$ imply a GDW between $\B$ and $\FF$ directly. $\B'$ is compressible between $B$ and $\FF$.}
\label{fig:GDWGDW}
\end{figure}

The above scenario is physically sound. On the one hand, the entire system in Fig. \ref{fig:GDWGDW}(a) is gapped, including the domains walls and $\B'$. In fact, one can view $\B'$ together with the two GDWs  $W_{\B\B'}$ and $W_{\B'\FF}$ as a generalized, $2$-dimensional GDW. Hence, one can imagine shrinking the width of $\B'$ to zero, in which limit the two gapped domain walls $W_{\B\B'}$ and $W_{\B'\FF}$ would become a single GDW, i.e., Fig. \ref{fig:GDWGDW}(b). On the other hand, as alluded to earlier, a GDW serves as a map of the anyons in one topological order to those in another. Such maps can certainly be concatenated into a single map. In purely bosonic cases, a GDW is represented by a rectangular matrix, and the above process of combining two GDWs is merely a matrix multiplication\cite{HungWan2014,Lan2015}. A GDW between a bTO and an fTO would also be represented by a matrix; however, in this paper we shall not address this for the following reason. In purely bosonic cases, a matrix representing a GDW between two bTOs is either the mass matrix of  a nonchiral RCFT or the branching matrix of certain vertex algebra embedding.\cite{HungWan2015a} As such, boson condensation corresponds to modular invariants of RCFTs. In the cases with GDWs between bTOs and fTOs, however, we do not yet know the role of modular invariants played here because not only the usual CFTs but also fermionic CFTs are involved. We thus would like to defer the investigation of how fermion condensation may corresponds to fermionic RCFT to future work.

There is another subtlety in the purely bosonic case as well: If there is a global symmetry on the relevant bTOs, i.e., if one is actually dealing with SETs, the compressibility of a bTO between two other bTOs may not be guaranteed because the global symmetry may also impose extra conditions on boson condensation. Similar subtlety may also arise in our current considerations.

One may also think of the GDWs between two fTOs $\FF$ and $\FF'$. Such GDWs can correspond to two kinds of anyon condensation. The first is boson condensation in $\FF$ that takes $\FF$ to $\FF'$. In this case, boson condensation follows the same rules as those of boson condensation in bTOs, while the super partners of the condensed self-bosons are treated as usual anyons. We shall see an example in Section \ref{subsec:su2}. The second is fermion condensation in $\FF$. The result is certainly also an fTO, say, $\FF'$. Via the folding trick along the GDW between $\FF$ and $\FF'$, seen in Fig. \ref{fig:gdwAndFoldingFFprime}, we obtain an equivalent configuration, $\FF\bx_{\FF_0}\FF'$, with a gapped boundary separating the vacuum. By the same token as in the remark below Fig. \ref{fig:gdwAndFolding}, such fermion condensation is equivalent to boson condensation in $\FF\bx_{\FF_0}\FF_0 = \FF$. Hence, boson condensation and fermion condensation in fTOs are actually the same thing.
\begin{figure}[h!]
\subfigure[GDW due to fermion condensation]{\label{subfig:gdwFFprime}
\gdw{\FF}{\FF'}
}
\subfigure[Folding trick]{\label{subfig:foldFFprime}
\gdwFold{\FF\bx_{\FF_0}\overline{\FF'}}{\FF_0}
}
\caption{(a) Two fTOs $\FF$ and $\FF'$ connected by a GDW via condensing some fermions in $\FF$. Via the folding trick, this picture is equivalent to (b) A non-prime fTO $\FF\bx_{\FF_0}\overline{\FF'}$.}
\label{fig:gdwAndFoldingFFprime}
\end{figure}

\section{examples}\label{sec:example}
In this section, we apply our principles of fermion condensation developed in previous sections to a few examples explicitly.

\subsection{$\B=\Z_2$-Toric Code}\label{subsec:Z2}
We first go through a simple example, the $\Z_2$ toric code. This is an Abelian bTO with four anyons, $1$, the trivial boson, $e,m$, which are self-bosons, and a self-fermion $\epsilon$. All anyons here are self-dual and simple currents. The nontrivial fusion rules are $e\x m=\epsilon$, $e\x \epsilon=m$. The $\Z_2$ toric code thus have a unique minimal fermion condensation $A^F_0=1\oplus \epsilon$. According to the rule of confinement listed in the previous section, because $e\x\epsilon=m$ and $\theta_e-\theta_m=0$, $e$ and $m$ will be confined and do not appear in the child fTO $\FF$, which in this example is simply $\FF_0=\{1,1^f\}$. Clearly, the $\epsilon$ becomes $1^f$ and no nontrivial anyons survive the $A^F_0$ condensation. In other words, there is a gapped boundary between the $\Z_2$ toric code and the vacuum, $e$ and $m$ cannot cross the boundary into the vacuum. In this example, since $D_\B=2$, and $D_{\FF_0}=\sqrt{2}$, the dimension formula \eqref{eq:dimRelPrime} is obviously satisfied.

\subsection{$\B=SU(2)_{4l-2}$}\label{subsec:su2}
We then look at a family of examples, which are specified by the representation categories of the affine Lie algebra $SU(2)_k$ for $k=4l-2$, $l\in\Z_+$. Boson condensation in this family of bTOs have been studied thoroughly\cite{HungWan2015a}, and now we study $A^F_0$ condensation in these bTOs.

In an $SU(2)_{4l-2}$, the anyons $a$ are labeled by integers, $a=0,1,2,\dots,4l-2$. The quantum dimensions and topological spins are respectively
\be\label{eq:su2qd}
\begin{aligned}
d_a&=\frac{\sin[(a+1)\pi/4l]}{\sin(\pi/4l)},\\
h_a&=\frac{a(a+2)}{16l}\pmod{1}.
\end{aligned}
\ee
Clearly, the anyon $a=4l-2$ for any $l$ is a simple current self-fermion, i.e., $d_{4l-2}\equiv 1$ and $h_{4l-2}=(2l-1)/2$. The fusion rule for any $a$ and $b$ is
\be\label{eq:su2fusion}
a\x b=c_{ab}+(c_{ab}+2)+\cdots + \min\{a+b, 8l-4-a-b\},
\ee
where $c_{ab}=|a-b|$. So, all anyons are self-dual. Hence, we have for any such bTO with a given $l$ a minimal fermion condensation $A^F_0=0\oplus (4l-2)$. Condensing $A^F_0$ results in an fTO $\FF_{SU(2)_{4l-2}}$. To find the anyons in $\FF_{SU(2)_{4l-2}}$ we need the unconfined anyons of $SU(2)_{4l-2}$ due the condensation. For any $a$, its fusion with the condensed self-fermion $4l-2$ yields exactly one anyon, $a \x (4l-2)= (4l-2-a)$, by the fusion rule \eqref{eq:su2fusion}. If $|h_a-h_{4l-2-a}|\not\in \Z_+/2$, $a$ and its super partner $4l-2-a$ will be confined. Using Eq. \eqref{eq:su2qd}, by elementary arithmetics, we obtain for each $l$ value an fTO:
\be
\FF_{SU(2)_{4l-2}}=\{a_n,a_n^f|a_n=2n,n=0,\dots,l-1\},
\ee
where $a_n^f=4l-2-a_n$, in terms of the $SU(2)_{4l-2}$ labels, is the super partner of $a_n$. The trivial boson is $a_0=0$, and trivial fermion is $a_0^f=4l-2$. . We emphasize that in our notation, the superscript $f$ in $a_n^f$ does not imply that $a_n^f$ is a fermion at all, as it only indicates that $a_n^f$ differs from $a_n$ by a fermion parity because $a_n^f=a_n\x a_0^f$. This is no more than a mere choice of convenience. The anyons in $\FF_{SU(2)_{4l-2}}$ fuse in exactly the same way as if they were $SU(2)_{4l-2}$ anyons.

Among this family of examples, let us focus on the case with $l=3$, i.e., $SU(2)_{10}$, to illustrate the Hierarchy Principle discussed in the previous section. To this end, we tabulate the topological data of $\FF_{SU(2)_{10}}$ as follows.
\begin{table}[h!]
\centering
\setlength\extrarowheight{5pt}
\resizebox{0.45\textwidth}{!}{%
\begin{tabular}{l|c|c|c|c|c|c|}
\hline
\multicolumn{1}{|l|}{$d_{a_n}$} & $1$ & $1$ & $1+\sqrt{3}$ & $1+\sqrt{3}$ & $2+\sqrt{3}$ & $2+\sqrt{3}$ \\ \hline
\multicolumn{1}{|l|}{$h_{a_n}$} & $0$ & $\tfrac{1}{2}$ & $\tfrac{1}{6}$ & $-\tfrac{1}{3}$ & $\tfrac{1}{2}$ & $0$ \\ [0.5ex] \hline
\multicolumn{1}{c|}{} & $0$ & $0^f$ & $2$ & $2^f$ & $4$ & $4^f$ \\ \hline
\multicolumn{1}{|l|}{$0$} & $0$ & $0^f$ & $2$ & $2^f$ & $4$ & $4^f$ \\ \hline
\multicolumn{1}{|l|}{$0^f$} & $0^f$ & $0$ & $2^f$ & $2$ & $4^f$ & $4$ \\ \hline
\multicolumn{1}{|l|}{$2$} & $2$ & $2^f$ & $0+2+4$ & $0^f+2^f+4^f$ & $2+4+4^f$ & $4+4^f+2^f$ \\ \hline
\multicolumn{1}{|l|}{$2^f$} & $2^f$ & $2$ & $0^f+2^f+4^f$ & $0+4$ & $2^f+4+4^f$ & $2+4+4^f$ \\ \hline
\multicolumn{1}{|l|}{$4$} & $4$ & $4^f$ & $2+4+4^f$ & $2^f+4+4^f$ & $0+2+2^f+4+4^f$ & $0^f+2+2^f+4+4^f$ \\ \hline
\multicolumn{1}{|l|}{$4^f$} & $4^f$ & $4$ & $4+4^f+2^f$ & $2+4+4^f$ & $0^f+2+2^f+4+4^f$ & $0+2+2^f+4+4^f$ \\ \hline
\end{tabular}%
}
\caption{Quantum dimensions, topological spins, and fusion rules of $\FF_{SU(2)_{10}}$, where $a_0$ and $a_0^f$ are renamed to $0$ and $0^f$.}
\label{tab:Fsu2_10}
\end{table}

The case with $l=3$ also illustrates the scenario shown in Fig. \ref{fig:GDWGDW}. We again begin with $\B_{SU(2)_{10}}$. This bTO contains a boson condensation $A^B=0\oplus 6$, which has been shown to break $\B_{SU(2)_{10}}$ to the $SO(5)_1$ topological order $\B_{SO(5)_1}=\{1,\sigma,\psi\}$, where $\psi$ is a self-dual simple current self-fermion. The bTO $\B_{SO(5)_1}$ is a sibling of $\B_{\rm{Ising}}$ and has a minimal fermion condensation $A_{\FF_0}=1\oplus\psi$, as mentioned earlier in the work. Condensing the $A_{\FF_0}$ breaks $\B_{SO(5)_1}$ into the trivial fTO $\FF_0$. 

This two-step condensation process can be combined into a single fermion condensation of $A^F$. First, the $\psi$ in here in $\B_{\rm{Ising}}$ descends from the anyons $4$ and $10$ of $\B_{SU(2)_{10}}$ through condensing the $A^B$. Thus, if one would combine the $A^B$ and $A^F_0$ here as a nontrivial fermion condensation in $\B_{SU(2)_{10}}$, one would have to invoke the condensation of $4$ and $10$, which is legitimate because this does not violate any of conditions of the ansatz. In fact, this two step condensation $(A^B=0\oplus 6)\orbx (A^F_0=1\oplus\psi)$ can be promoted to the nontrivial fermion condensation $A^F=0\oplus 6\oplus 4\oplus 10\in SU(2)_{10}$, whose condensation directly breaks $SU(2)_{10}$ into $\FF_0$. The discussion above simply manifests the hierarchy principle of condensing this $A^F$.

Yet, an alternative route exists to achieve the final trivial fTO $\FF_0$. The $4^f$ in $\FF_{SU(2)_{10}}$ is a self-boson. It is actually the anyon $6$ in $SU(2)_{10}$ before condensing the $A^F_0$. In $SU(2)_{10}$, the anyon $6$ is allowed to condense individually, as studied in Ref.\cite{Bais2009,HungWan2015a}. In $\FF_{SU(2)_{10}}$, $4^f$ can condense as well, and its condensation follows the same rules of boson condensation in bTOs. The result is again the trivial fTO $\FF_0$. These two steps of condensation can also be combined into the fermion condensation $A^F=0\oplus 6\oplus 4\oplus 10\in SU(2)_{10}$. What we just did is condensing $A^F_0=0\oplus 10$ first, which led to $\FF_{SU(2)_{10}}$, followed by condensing ${A^B}'=1\oplus 4^f\in \FF_{SU(2)_{10}}$ (note that $4^f$ is a boson). Therefore, we have the following commutative diagram of anyon condensation.
\be\nonumber
\begin{tikzcd}
SU(2)_{10} \arrow{r}{0\oplus 6} \arrow{rd}{A^F} \arrow[swap]{d}{0\oplus 10} & SO(5)_1 \arrow{d}{1\oplus\psi} \\
\FF_{SU(2)_{10}}  \arrow[swap]{r}{1\oplus 4^f} & \FF_0
\end{tikzcd}.
\ee

Seen from the commutative diagram above, the fermion condensation $A^F=0\oplus 6\oplus 4\oplus 10$ here also admits a different hierarchical decomposition, namely the minimal fermion condensation $A^F_0=0\oplus 10$ followed by the boson condensation ${A^B}'=1\oplus 4^f$. This is because minimal fermion condensation involves merely a simple current fermion, whose condensation does not cause any boson condensation in general. Hence,one can first condense the $A^F_0$ and then do the residual boson condensation in the resulted fTO. Clearly, such a decomposition is possible only if certain fermion condensation directly contains some $A^F_0$. One will see a counterexample in the next example.

Finally, as we mentioned previously, fermion condensation in bTOs is equivalent to boson condensation in fTOs, via the folding trick. In the current case, the $A^F=0\oplus 6\oplus 4\oplus 10$ condensation in $\B_{SU(2)_{10}}$ can be equivalently understood as condensing certain bosons in the fTO  $\B_{SU(2)_{10}}\bx \FF_0$, where anyons are again labeld by $0, 0^f, 1,1^f, \dots, 10, 10^f$. If we condense $A^B=0\oplus 6\oplus 4^f\oplus 10^f$, we end up the trivial fTO $\FF_0$.

\subsection{$\B=D[D_3]$}\label{subsec:DD3}
This is another example with nonsimple current self-fermion condensation, which certainly also induces self-boson condensation. The parent bTO is the quantum double $D[D_3]$ of the dihedral group $D_3$ that is isomorphic to the permutation group $S_3$ of three elements. We need not to get into the detail of how the mathematical structure of the quantum double $D[D_3]$ is constructed and how the anyon content is obtained. For our purpose, we simply tabulate in Table \ref{tab:DD3} all the eight anyons of $D[D_3]$ and their properties and work on the condensation from there.
\begin{table}[h!]
\setlength\extrarowheight{5pt}
\centering
\resizebox{0.25\textwidth}{!}{%
\begin{tabular}{l|cccccc}
 & $1$ & $e_1$ & $e_2$ & $m^{l=0,1,2}_2$ & $m^+_3$ & $m^-_3$ \\ \hline
$d$ & $1$ & $1$ & $2$ & $2$ & $3$ & $3$ \\
$h$ & $0$ & $0$ & $0$ & $\tfrac{l}{3}$ & $0$ & $\tfrac{1}{2}$
\end{tabular}%
}
\caption{The eight anyons of the quantum double $D[D_3]$. The total quantum dimension is $D_{D[D_3]}=6$.}
\label{tab:DD3}
\end{table}

Because all the quantum dimensions are integers anyway in a quantum double of any finite group, for mnemonics, the subscript (an integer) in each anyon label explicitly indicates the quantum dimension of the anyon. The anyon $m^-_3$ is a self-fermion, and to see whether it can condense, we need to look at the fusion rules as follows. We shall neglect the trivial fusion rule between the trivial boson and other anyons.
\begin{table}[h!]
\setlength\extrarowheight{5pt}
\centering
\resizebox{0.45\textwidth}{!}{%
\begin{tabular}{l}
\toprule
$ e_1\x e_1=1$\hspace{2ex}    $e_1\x e_2=e_2$ \\
$e_2\x e_2= 1+e_1+e_2$ \\[0.5ex]\hline
$e_1\x m^l_2=m^l_2$ \hspace{2ex}    $e_1\x m^+_3=m^-_3$\hspace{2ex} $e_1\x m^-_3=m^+_3$ \\
$e_2\x m^l_2=m^j_2+m^k_2,\ \ \ l\neq j\neq k$ \\
$e_2\x m^+_3=m^+_3+m^-_3$\hspace{2ex} $e_2\x m^-_3=m^+_3+m^-_3$ \\[0.5ex] \hline
$m^l_2\x m^l_2=1+e_1+m^l_2$\hspace{2ex}  $m^l_2\x m^j_2=e_2+m^k_2,\ \ \ l\neq j\neq k$ \\
$m^l_2\x m^+_3=m^+_3+m^-_3$ \hspace{2.5ex}     $m^l_2\x m^-_3=m^+_3+m^-_3$ \\[0.5ex] \hline
$m^+_3\x m^+_3=1+e_2+\sum_{l=0,1,2} m^l_2$ \\
$m^+_3\x m^-_3 = e_1+e_2+\sum_{l=0,1,2} m^l_2$ \\
$m^-_3\x m^-_3 = 1+e_2+\sum_{l=0,1,2} m^l_2$ \\ [1ex] \hline
\end{tabular}%
}
\caption{Fusion Rules of $D[D_3]$.\cite{Propitius1995,Bais2009a}}
\label{tab:FusionDD3}
\end{table}

The self-fermion $m^-_3$ is a nonsimple current, as it has quantum dimension $3$; hence, according to Lemma \ref{lem:minFC}, it cannot condense individually but has to induce certain boson condensation. Staring at the last row of Table \ref{tab:FusionDD3}, the self fusion of $m^-_3$, since $e^2$ and $m^0_2$ are both self-bosons, $m^-_3$ is self mutual-local via $e_2$ and $m^0_2$. Thus condensing $m^-_3$ might force $e_2$ and/or $m^0_2$ to condense too. But in the fourth row of Table \ref{tab:FusionDD3}, the fusion between $e_2$ and $m^0_2$ says that they are not partially mutual local and thus cannot condense simultaneously.

For our purposes, we choose $e_2$ to condense along with $m^-_3$. The second row of Table \ref{tab:FusionDD3} implies that $e_2$'s condensing might result in $e_1$'s condensing. Nevertheless, because $e_1\x m^-_3=m^+_3$, $e_1$ and $m^-_3$ are not partially mutual local and thus cannot condense together. We then propose the fermion condensation---a fermionic Frobenius algebra---$A^F_{D[D_3]}=1\oplus e_2\oplus m^-_3$. This condensation $A^F_{D[D_3]}$ satisfies the ansatz, all conditions of which are obviously fulfilled except the trivial associativity. It is highly nontrivial and requires the $F$-symbols to show that $A^F_{D[D_3]}$ is trivially associative. So, we would not do the proof in this paper. On the other hand, $A^F_{D[D_3]}$ clearly satisfies the necessary condition of trivial associatively, namely the condition in Lemma \ref{lem:associativity}. Therefore, we shall just proceed to condense $A^F_{D[D_3]}$, which is a perfect example to exhibit the hierarchy principle of fermion condensation.

The fermionic Frobenius algebra $A^F_{D[D_3]}$ has a bosonic subalgebra $A^B_{D[D_3]}=1\oplus e_2$. Then, let us first condense $A^B_{D[D_3]}$. This boson condensation has been studied in detail in Ref.\cite{Bais2009a}. After this boson condensation, the child bTO $\B'$ of $\B_{D[D_3]}$ is the $\Z_2$ toric code, which contains a simple current self-fermion $\epsilon$ (see Section \ref{subsec:Z2}). This self-fermion $\epsilon$ is in fact a descendent of the $m^-_3\in A^F_{D[D_3]}$ that splits into two parts due to the condensation of $e_2$. More explicitly, with condensing $e_2$, we have the splitting $m^-_3\rightarrow m^-_{3,1}+m^-_{3,2}$. To understand this splitting, we can apply the rules of boson condensation reviewed in Section \ref{subsubsec:revBC} and the fusion rules in Table \ref{tab:FusionDD3}:
\be\nonumber
\begin{aligned}
& m^-_3\x A^B_{D[D_3]}= (m^-_3 ) \oplus (m^+_3\oplus m^-_3),\\
& m^+_3\x A^B_{D[D_3]}= (m^+_3 ) \oplus (m^+_3\oplus m^-_3).
\end{aligned}
\ee
It is then not hard to see that in the child bTO (i.e., the $\Z_2$ toric code) after the $A^B_{D[D_3]}$ condensation, $m^-_3$ and $m^+_3\oplus m^-_3$ are both simple objects that are not of the form of $F(a)$ for any $a\in D[D_3]$. We thus do the redefinition $m^-_{3,1}= m^-_3|_{\scalebox{0.5}{$\B'$}}$ and $m^-_{3,2}= (m^-_3\oplus m^+_3)_{\scalebox{0.5}{$\B'$}}$, where the subscript $\B'$ is clearly a restriction. We can obtain the quantum dimensions of the two parts of $m^-_3$ as
\be\nonumber
\begin{aligned}
& d_{m^-_{3,1}} = \frac{\dim_\B m^-_3}{\dim A^B_{D[D_3]}} = \frac{d_{m^-_3}}{1+d_{e_2}}=1,\\
& d_{m^-_{3,2}} = \frac{\dim_\B (m^-_3\oplus m^+_3)}{\dim A^B_{D[D_3]}} = \frac{d_{m^-_3}+d_{m^+_3}}{1+d_{e_2}}=2,
\end{aligned}
\ee
where the conservation of quantum dimension through splitting is evident: $d_{m^-_3}=d_{m^-_{3,1}}+d_{m^-_{3,2}}$. Since $m^-_{3,2}$ descends from the identification of $m^-_3$ and $m^+_3$ that have topological spins different not by an integer, $m^-_{3,2}$ must be confined after condensing $A^B_{D[D_3]}$. But $m^-_{3,1}$ descends solely from $m^-_3$, it is thus unconfined and inherits the topological spin $1/2$ of $m^-_3$ and become a self-fermion in $\B'$, namely the $\epsilon$ in the $\Z_2$ toric code. We are not concerned of the rest of the $A^B_{D[D_3]}$ condensation, which can be found in Ref.\cite{Bais2009a}, but continue to finish the hierarchical condensation of $A^F_{D[D_3]}$.

Now that we have $\B'$ being the $\Z_2$ toric code, we naturally have the decomposition $A^F_{D[D_3]}=A^B_{D[D_3]}\orbx A^F_0$, where $A^F_0=\{1,\epsilon\}$. Hence, the next and final step is to condense the $\epsilon$ in the $\Z_2$ toric code. This minimal fermion condensation is discussed in Section \ref{subsec:Z2}, and the result is the trivial fTO $\FF_0$. Our dimension formula \eqref{eq:dimRel} is verified:
\be
\sqrt{2}\frac{D_{D[D_3]}}{\dim A^F_{D[D_3]}}=\sqrt{2}\frac{6}{1+2+3}=\sqrt{2}=D_{\FF_0}.
\ee

\section{Discussion}\label{sec:Disc}
Following the ansatz of fermion condensation we lay down, we obtain a hierarchy principle of fermion condensation, which permits decomposing arbitrary fermion condensation into certain boson condensation followed by minimal fermion condensation. The rules of minimal fermion condensation we then find and the previously known rules of boson condensation combine into a full prescription of performing fermion condensation in bTOs. 
Fermion condensation, now in a more general sense by including boson condensation as special cases, naturally corresponds to GDWs between topological orders, both bTOs and fTOs. Our results are supported by the explicit examples worked out in the section above.

Boson condensation in a parent bTO gives rise to a linear mapping between the Hilbert spaces of the parent bTO and that of the child bTO obtained from the parent one via the condensation, which are studied in detail in Ref.\cite{Eliens2013,Gu2014a}. This mapping enables one to express the topological quantities, such as the modular $S$ and $T$ matrices, of the child bTO in terms of those of the parent bTO\cite{Eliens2013}. This mapping also offers a toolkit for extracting the global symmetry action on the child bTO, which is an effect of the confined anyons due to the boson condensation involved. In this sense, the child bTO is in fact an SET order. Fermion condensation in a bTO also leads to global symmetries on the child fTO of the bTO. On the one hands, the boson condensation contained in the fermion condensation would result in a usual global symmetry on the child fTO as it does in the case with pure boson condensation. Likewise, the minimal fermion condensation after the boson condensation would yield the fermion parity symmetry on the child fTO. It is important and reasonable to expect that fermion condensation can also lead to a mapping between the Hilbert space of a parent bTO and that of the child fTO of the parent one. Such a mapping is plausibly projective because the vacuum of an fTO is not trivial but consists of both trivial bosons and trivial fermions and thus has an internal space. In Appendix \ref{app:trivialAssoc}, we slightly touch upon the search of such mapping by restricting to the vacua of the parent bTO and the child fTO only. By doing so, we prove Lemma \ref{lem:associativity}. Nevertheless, because of the potential subtleties caused by the trivial fermions in an fTO, we are still looking for a rigorous and appropriate description of the full Hilbert space of an fTO. We shall report our progress along this direction elsewhere.

In Ref.\cite{HungWan2015a}, it is found that boson condensation is in one-to-one correspondence with vertex operator algebra embedding and hence with modular invariants in RCFTs. It would be interesting to see if fermion condensation would have a similar correspondence with the modular invariants in fermionic RCFTs.

In the end, we would also like to touch upon the possibility of condensing anyons more exotic than self-fermions, e.g., semions, etc. Such discourses are however way beyond not only the focal range of the current work but also our rudimental understanding of the consequences. Hence, we would rather note down our preliminary thoughts of general anyon condensation in Appendix \ref{app:genAnyonCond}.

\acknowledgements{The authors appreciate Davide Gaiotto for his deep insights into the problem of fTOs, helpful discussions, and comments on the manuscript. YW is grateful to Jurgen Fuchs for helpful answers to his questions regarding Frobenius algebras and helpful comments on the manuscript. YW also thanks Yuting Hu for helpful discussions. YW is supported by the John Templeton foundation No. 39901. CW thanks the Aspen Center for Physics for hospitality, where part of this work is completed. The Aspen Center for Physics is supported by National Science Foundation grant PHY-1066293. This research was supported in part by Perimeter Institute for Theoretical Physics. Research at Perimeter Institute is supported by the Government of Canada through the Department of Innovation, Science and Economic Development Canada and by the Province of Ontario through the Ministry of Research, Innovation and Science.}

\appendix
\section{Review of Frobenius algebra}\label{app:Frobenius}
We briefly review in physical terms the concept of twist-free commutative separable Frobenius algebra (CSFA) of a UMTC $\C$. This mathematical structure has been used for classifying the quantum subgroups of quantum groups, the vertex operator algebra embedding in $2$-dimensional rational conformal field theories, and domain walls between $3$-dimensional topological field theories.\cite{Kirillov2002} Thorough studies of  twist-free CSFAs can be found in Ref.\cite{Kirillov2002,Fuchs2002,Kong2013} for example.

{\bf Frobenius algebra\footnote{A Frobenius algebra is not only an algebra but also a coalgebra. In this paper, however, we does not need the coalgebra aspect of a Frobenius algebra and thus do not bring it up}.}\,\,\,  A UMTC $\C$ admits a special type of objects---Frobenius algebra objects. A Frobenius algebra $A$ in this context is generally a direct sum of certain simple objects of $\C$. For a bTO described by $\C$, the simple objects are the distinct elementary anyon types. This object $A$ is a Frobenius algebra because 1) it is endowed with a product $\mu:A\ox A\rightarrow A$ by the fusion of the simple objects of $\C$, 2) an inclusion $\iota_A:1_\C\hookrightarrow A$, where $1_\C$ is the unit object or vacuum of $\C$, such that $1_\C$ is also the unit of $A$, 3) the product $\mu$ is associative, and 4) haploid: the unit is unique, namely $\dim \Hom_\C(1_\C,A)=1$. A commutative Frobenius algebra is one whose product $\mu$ commute with the braiding of $A$. This means $\mu\circ R_{AA}=\mu$, where $R_{AA}$ is the $R$ matrix of $A$ in $\C$. This commutativity is physically sound in the case with boson condensation because $A$ is going to become the new vacuum when it condenses. In the case of fermion condensation, since $R_{ff}^1=-1$ for any condensed fermion $f$, the commutativity may cease to hold.

One may formally write $A=1\oplus\Upsilon$, where $\Upsilon$ is the direct sum of the nontrivial simple objects of $\C$ that comprise $A$. An $A$ describing anyon condensation is necessarily self-dual, also called {\bf \textit{rigid}} in Kirillov\cite{Kirillov2002}. This means mathematically there is a non-degenerate projection from $A\ox A$ to $1_\C$. Physically, viewed as a composite anyon in $\C$, $A$ is the anti-anyon of itself, consistent with that $A$ would become the new vacuum after its condensation.

{\bf Representation category Rep$A$}. There exists a representation category $\rep A$ over $A$. One can define the twist $\theta_A$ of $A$ as the self-statistical angle of $A$, as $A$ is an object in $\C$. If $A$ is {\bf \textit{twist-free}}, $\theta_A=\id_A$, then $\rep A$ is a tensor category. The twist-free condition is equivalent to one that any simple object in $A$ is a self-boson. Physically, this means that $A$ specifies certain boson condensation, and after its condensation, $A$ would become the new trivial boson, or vacuum. For $\rep A$ to be also semisimple, $A$ needs to be {\bf \textit{separable}}, which gives rise to well-defined simple objects in $\rep A$. The tensor products of the simple objects of $\rep A$ can be written as direct sums of the simple ones. These simple objects are the superselection sectors to be identified as anyons or defects after condensing $A$. A trivial example of a twist-free CSFA is $A=1_\C$ in any $\C$, meaning no actual anyon condensation.

In general, $\rep A$ is not braided but has a braided subcategory $\rep_0 A$ that consists of the objects in $\rep A$ satisfying the criterion
\be
\rep_0 A=\{X\in\T|(A\otimes_A X)R_{XA}R_{AX}=A\otimes_A X\},
\ee
with the fusion $\ox_A$ defined with respect to $A$. The details can be found in Ref.\cite{Kirillov2002,Frohlich2006} and not repeated here.

Apparently, a twist-free CSFA cannot describe any fermion condensation. In order to do so, it seems that the twist-free condition and commutativity would be relaxed. Unfortunately, the mathematically rigorous consequences of relaxing these two conditions are yet unclear, which deserves further investigation.
\section{A Necessary Condition of the Trivial Associativity of Condensation}\label{app:trivialAssoc}
We show how we obtain Lemma \ref{lem:associativity}, namely a necessary condition of the trivial associativity condition in our ansatz of fermion condensation.

In cases of boson condensation, the Hilbert space of a parent bTO $\B$ can be linearly mapped to that of the child phase $\B'$ of $\B$ via certain boson condensation, and vice versa. The Hilbert space of $\B$ is a fusion space, decomposed into subspaces, each of which is specified by the number of anyon excitations and the types of anyons. All such subspaces can be built on top of the basic fusion space $\Hil_3$, which involves three anyons only. An $\Hil_3$ is spanned by the basis vectors $\ket{\Psi^{ab}_c}$,\footnote{Here we assume that $N^{ab}_c=1$ for simplicity. Generalization to cases with $N^{ab}_c\neq 1$ is straightforward.\cite{Gu2014a}} where $a,b,c\in\B$ and $\Psi^{ab}_c$ the spatial wavefunction of this state. Note that the state vector of a single anyon excitation $a$ is a special basis vector $\ket{\Psi^{a1}_a}=\ket{\Psi^{1a}_a}\in\Hil_3$. Since, attaching a trivial boson $1$ to an anyon is somewhat tautological, we may also denote a single anyon state by $\ket{\Psi^a}$ whenever it is necessary to do so. Canonically, the basis vector $\ket{\Psi^{ab}_c}$ refers to the particular fusion channel $c\in a\x b$. The fusion space $\Hil_3$ has a dual space $\Hil_3^*=\{\ket{\Psi^c_{ab}}|a,b,c\in\B\}$, where $\Psi^c_{ab}={\Psi^{\bar a\bar b}_{\bar c}}^*$ is the wavefunction relation. A $\ket{\Psi^c_{ab}}$ may be viewed as $a$ and $b$ fuse into $c$.

Here is an important remark. In the definition of the dual space $\Hil_3$, we still use kets rather than bras. So, the duality lies in the wavefunction normalization
\be\label{eq:PsiNorm}
\Psi^c_{\dot a\dot b}\Psi^{\dot a\dot b}_{c'}=\delta_{cc'}\sqrt{d_a d_b/d_c}\Psi^{c1}_c,
\ee
where we used $\dot x$ to indicate the internal anyons. In other words, we write all the state vectors in terms of the corresponding wavefunctions. We take such an unusual convention to avoid the potential ambiguity\footnote{Private discussion with Davide Gaiotto.} of representing a transparent fermion by a line in fTOs. As such, we can build the states in a Hilbert space of four anyons, e.g., $\ket{\Psi^a_{b\dot m}\Psi^{\dot m d}_c}=\ket{\Psi^a_{b\dot m}} \ox \ket{\Psi^{\dot m d}_c}$, without summing over the internal fusion channel $m$. Note that the $\dot{}$ in $\dot m$ is not part of the index labeling the anyon but merely indicates that $m$ is internal. The associativity of fusion reads
\be\label{eq:fusionAssoc}
\ket{\Psi^a_{b\dot m}\Psi^{\dot m d}_c}=\sum_{m'}[F^{ad}_{bc}]^m_{m'}\ket{\Psi^{ad}_{\dot m'}\Psi^{\dot m'}_{bc}},
\ee
where the complex coefficients are the $F$-symbols. The Definition and properties of $F$-symbols can be easily found in any standard introduction to topological orders or tensory categories. Here we need only the usual normalization of $F$-symbols
\be\label{eq:Fnorm}
[F^{ab}_{ab}]^0_c=\sqrt{\frac{d_c}{d_a d_b}}N^c_{ab}.
\ee
we expect that such a description of Hilbert spaces still applies to fTOs. Nonetheless, since we will restrict ourselves to the vacuum of any fTO in this appendix, we may leave any subtleties of the Hilbert space of an fTO for future studies.

In cases with boson condensation, the Hilbert spaces of $\B$ and $\B'$ are related by a linear map, which, when boiled down to $\Hil^3$ reads
\be\label{eq:vlc}
\ket{\Psi^{\alpha\beta}_\gamma}=\sum_{a,b,c\in\B}\vlc{\alpha}{\beta}{\gamma}{a}{b}{c}\sum_{a,b,c}\ket{\Psi^{ab}_c}, \ \ \ \forall \alpha, \beta,\gamma\in\B',
\ee
where $\vlc{\alpha}{\beta}{\gamma}{a}{b}{c}$ are complex coefficients, which are called vertex lifting coefficients (VLC)\cite{Eliens2013,Gu2014a} because we \textquotedblleft lift" a fusion vertex $\Psi^{\alpha\beta}_{\gamma}$ of the child TO to fusion vertices $\Psi^{ab}_c$ of the parent TO by the linear map. The VLCs satisfy certain consistency conditions.\cite{Eliens2013,Gu2014a} But we do not need any such details here.

Back in cases with fermion condensation, because fermionic vacua are nontrivial, we would expect the map between a parent bTO $\B$ and its child fTO $\FF$ to be at least projective rather than linear. We shall report detailed studies of the map elsewhere. In this appendix, our goal is to find a necessary condition of the trivial associativity of any fermion condensation $A^F\in\B$, we need only to focus on the relation between the vacuum of $\FF$ and $A^F$. Hence, restricted to the vacuum space of $\FF$, its relation with the Hilbert space of $\B$ restricted to $A^F$ would still be linear. As such, we can rewrite the VLC equation \eqref{eq:vlc} as
\be\label{eq:vlcVac}
\ket{\Psi^{\alpha\beta}_\gamma}=\sum_{a,b,c\in\A^F}\phi^{ab}_c\sum_{a,b,c}\ket{\Psi^{ab}_c}, \ \ \ \forall \alpha, \beta,\gamma\in\{1,1^f\}\subseteq\FF,
\ee
where $\phi^{ab}_c$ are the VLCs simplified when restricted to the vacuum of $\FF$. Note that any vacuum state $\ket{\Psi^{\alpha\beta}_\gamma}$ in the vacuum of $\FF$ has even fermion parity because $1\x 1^f=1^f$ and $1^f\x 1^f=1$. Because $d_1=d_{1^f}=1$, the normalization of vacuum wavefunctions, similar to Eq. \eqref{eq:PsiNorm}, is
\be\label{eq:vacPsiNorm}
\begin{aligned}
& \Psi^1_{\dot 1\dot 1}\Psi^{\dot 1 \dot 1}_1=\Psi{^1},\\
& \Psi^1_{\dot 1^f\dot 1^f}\Psi^{\dot 1^f \dot 1^f}_1=\Psi^1,\\
& \Psi^{1^f}_{\dot 1^f\dot 1}\Psi^{\dot 1^f \dot 1}_{1^f}=\Psi^{1^f},
\end{aligned}
\ee
where we emphasize on the one-particle states by $\Psi^1$ and $\Psi^{1^f}$. Now lift both sides of the equations in \eqref{eq:vacPsiNorm} following the defining equation \eqref{eq:vlcVac}, we obtain the following constraints.
\be\label{eq:phiConstraint}
\phi^{11}_1=\phi^{a1}_a=\phi^{1a}_a=\phi^{a\bar a}_1=1,\quad\forall a\in A^F.
\ee

We are now ready to look at the trivial associativity of $A^F$. As explained in Section \ref{subsec:FandGDW}, the trivial associativity of $A^F$ means that after $A^F$ condenses and becomes a fermionic vacuum, the $F$-symbols relating the vacuum states are an identity. In particular, Lemma \ref{lem:associativity} follows from the following two cases
\begin{align}
& \ket{\Psi^1_{1\dot 1}\Psi^{\dot 1 1}_1}=\ket{\Psi^{11}_{\dot 1}\Psi^{\dot 1}_{11}},\label{eq:all1}\\
& \ket{\Psi^{1^f}_{1^f\dot 1}\Psi^{\dot 1 1^f}_{1^f}}=\ket{\Psi^{1^f1^f}_{\dot 1}\Psi^{\dot 1}_{1^f1^f}}.\label{eq:all1f}
\end{align}
where the dotted ones are the internal trivial bosons/fermions. Let us focus on Eq. \eqref{eq:all1} first and lift its both sides. We have
\be\nonumber
\begin{aligned}
& \sum_{a,b,c,d,m\in A^F} {\phi^{a \dot m}_b}^*\phi^{\dot m d}_c \ket{\Psi^a_{b\dot m}\Psi^{\dot md}_c}\\
=& \sum_{a',b',c',d',m'\in A^F}|\phi^{a'b'}_{\dot m'}|^2 \ket{\Psi^{a'd'}_{\dot m'}\Psi^{\dot m'}_{b'c'}},
\end{aligned}
\ee
which holds term by term. Boths sides of the equation above now refers to the parent bTO $\B$; hence, applying the usual associativity \eqref{eq:fusionAssoc} to the LHS leads to
\be\nonumber
\begin{aligned}
& \sum_{a,b,c,d,m\in A^F}\sum_{m''\in A^F} [F^{ad}_{bc}]^m_{m''}{\phi^{a \dot m}_b}^*\phi^{\dot m d}_c \ket{\Psi^{ad}_{\dot m''}\Psi^{\dot m''}_{bc}}\\
=& \sum_{a',b',c',d',m'\in A^F}|\phi^{a'b'}_{\dot m'}|^2 \ket{\Psi^{a'd'}_{\dot m'}\Psi^{\dot m'}_{b'c'}},
\end{aligned}
\ee
where the $F$-symbols are those of $\B$. Now setting in the above $a=b=a'=b'=x$, $c=d=c'=d'=\bar x$, and $m'=m''=1_\B$, we obtain
\be\label{eq:trivialAssoc}
\sum_{m\in A^F}[F^{x\bar x}_{x\bar x}]^m_1{\phi^{x \dot m}_x}^*\phi^{\dot m \bar x}_{\bar x} \ket{\Psi^{x\bar x}_{\dot 1}\Psi^{\dot 1}_{x\bar x}}
=|\phi^{x\bar x}_{\dot 1}|^2 \ket{\Psi^{x\bar x}_{\dot 1}\Psi^{\dot 1}_{x\bar x}},
\ee
where the subscript in $1_\B$ is omitted for simplicity. Here comes the crucial step in proving Lemma \ref{lem:associativity}. Let us assume that $x$ is a nonsimple current, i.e., $d_x>1$, and $x$ and $\bar x$ do not produce any fusion channel that is also in $A^F$ except the trivial boson $1$. Then the summation in Eq. \eqref{eq:trivialAssoc} is gone, i.e., $m=1_\B$ only, and together with the constraints \eqref{eq:phiConstraint}, we would have
\be
[F^{x\bar x}_{x\bar x}]^1_1 \ket{\Psi^{x\bar x}_{\dot 1}\Psi^{\dot 1}_{x\bar x}}
= \ket{\Psi^{x\bar x}_{\dot 1}\Psi^{\dot 1}_{x\bar x}},
\ee
But by the $F$-symbol normalization \eqref{eq:Fnorm},
\be
[F^{x\bar x}_{x\bar x}]^1_1 \ket{\Psi^{x\bar x}_{\dot 1}\Psi^{\dot 1}_{x\bar x}}
=
\frac{1}{d_x} \ket{\Psi^{x\bar x}_{\dot 1}\Psi^{\dot 1}_{x\bar x}}
\neq \ket{\Psi^{x\bar x}_{\dot 1}\Psi^{\dot 1}_{x\bar x}},
\ee
obviously violating the trivial-associativity. Thus, the assumption is wrong. The fusion of $x$ and $\bar x$ must contain at least one nontrivial anyon that also condenses in order that the sum in the LHS of Eq. \eqref{eq:trivialAssoc} can turn out to be unity. One can do similar analysis for Eq. \eqref{eq:all1f} and obtain the same result. Therefore, we can conclude the validity of Lemma \ref{lem:associativity}.

A remark is that in the definition of fermion condensation in our fundamental ansatz, a condensable boson or fermion may have a multiplicity, i.e., it may appear multiple times in the Frobenius algebra describing the condensation. In our discussion in this appendix, we assume for simplicity all multiplicities being one. But this assumption causes no loss of generality because one can simply treat all occurrences of an anyon, had it have a nonunit multiplicity, as virtually distinct anyons with unit multiplicity.

The trivial associativity in our ansatz of fermion condensation may have more implications but we shall leave them for future work.

\section{Some preliminary thoughts of general anyon condensation}\label{app:genAnyonCond}
One may ask the question: is fermion condensation the most general anyon condensation? Certainly nature has not displayed any physical system whose fundamental degrees of freedom are semions or more exotically, anyons. It is however likely that certain highly nontrivial FQHS may be interpreted as a result of condensing anyons in less nontrivial FQHS (possibly multi-layer)\footnote{Private conversation with Xiao-Gang Wen}. Moreover, consider the idea of anyon superconductivity\cite{Laughlin1988,Laughlin1988a,Fetter1989,CHEN1989}, in which certain anyons may group into a composite anyon that behaves like a self-boson and can condense. It is in principle plausible to stack layers of trivial anyonic gases/liquids to certain bTO and condense a self-bosonic composite anyon made of certain anyon in the bTO and appropriate trivial anyons in the other layers. Neglecting the other layers but focusing on the bTO only, one can then study the principles and consequences of condensing the nontrivial anyon in the bTO, in a fashion similar to fermion condensation. This is in fact a justification via the folding trick, similar to that for boson condensation in Fig. \ref{fig:BgdwAndFolding}. To date, we are not sure about the physical barriers of condensing arbitrary anyons or the existence of fundamentally anyonic systems. But in theory or mathematically, there seems no a priori  an obstruction preventing one from studying general anyon condensation. Topological orders may be a playground for condensing generic anyons, such that at the effective level after the condensation, the vacuum of the system becomes anyonic, namely an anyonic liquid. To cater to anyonic vacua, we would have to modify the notion of mutual locality.

For example, imagine we had a semionic vacuum made of trivial semions $1^s$, with $\theta_{1^s}=\ii$; however, a semionic vacuum would automatically contain trivial fermions and trivial bosons too, as a fermionic vacuum contains both trivial fermions and trivial bosons. Since, $M^1_{1^s 1^s}=-1$, to account for a semionic vacuum, for a condensable anyon $a$ in a bTO, we would impose the mutual locality $M^1_{a\bar a}=\pm 1$. In a similar fashion, we can allow more exotic anyons to condense in a bTO by further relaxing the notion of mutual locality. This logic is justified because the topological sectors in a topological order may only be well-defined up to the vacuum structure. If the vacuum is fermionic, semionic, etc, the topological spin of a topological sector may be defined only up to fermion parity, semion type, etc. It turns out that we need only to relax the condition on mutual-locality in our fundamental ansatz of fermion condensation to accommodate general anyon condensation. Let us write this down.

\begin{ansatz*}
General anyon condensation of order $N$ in a single-layer bTO $\B$ is a composite object $A=n_a a\oplus n_b b\oplus\cdots\in \B$, where $a,b,\dots$ are anyons in $\B$ and $n_a$ the multiplicity (number of occurrences) of $a$ in $A$, that must meet the following conditions.
\begin{itemize}
\item[1.] Unit: $\B$'s vacuum---trivial boson---$1\in A$, $n_a=1$.
\item[2.] Self-dual: For any $a\in A$, the anti-anyon $\bar a\in A$.
\item[3.] Self-locality: For any $a\in A$, $(M^1_{a\bar a})^N=1$.
\item[4.] Closure: For any $a,b\in A$, there exists at least a $c\in a\x b$, such that $(M^c_{ab})^N=1$ and $c\in A$.
\item[4.] Mutual-locality: For any $a,b\in A$, if $c\in A$ and $c\in a\x b$, $a$ and $b$ are mutual-local via $c$.
\item[5.] Trivial associativity: $(A\x A)\x A=A\x(A\x A)$.
\end{itemize}
\end{ansatz*}

We leave any further studies on the consequences of this ansatz to future work.

\bibliographystyle{apsrev}
\bibliography{StringNet}
\end{document}